\documentclass[11pt,a4paper,english,twoside]{article}

\usepackage{a4wide}
\usepackage{amssymb, amsmath, amsthm}
\usepackage{graphicx}
\usepackage{subcaption}
\usepackage[all]{xy}
\usepackage{enumerate}
\usepackage[pdftex,hyperref,svgnames]{xcolor}
\usepackage[pdftex,colorlinks=true,
pdfstartview=FitV,
pdfnewwindow=true,
linktoc = page,
linkcolor= blue,
citecolor= red,
urlcolor= blue,
hyperindex=true,
hyperfigures=false]{hyperref}
\hypersetup{linktocpage}
\usepackage{dsfont}
\usepackage{empheq}
\usepackage{cite}
\usepackage{float}
\usepackage{cancel}
\usepackage{relsize}
\usepackage{soul}
\usepackage{enumitem}
\usepackage{hhline}
\usepackage{multirow}

\newcommand{\beq}{\begin{equation}}
\newcommand{\eeq}{\end{equation}}
\def\bea#1\eea{\begin{align}#1\end{align}}
\def\beal#1\eeal{\begin{subequations}\begin{align}#1\end{align}\end{subequations}}
\newcommand{\nn}{\nonumber}
\newcommand{\w}{\wedge}
\newcommand{\R}{\mathcal{R}}

\def\del {\partial}
\def\d {{\rm d}}

\newcommand{\ssum}{{\, \textstyle\sum\, }}

\newtheorem*{lemma}{Lemma 1}
\newtheorem*{lemma2}{Lemma 2}

\begin{document}
\numberwithin{equation}{section}

\begin{titlepage}

\begin{center}

\phantom{DRAFT}

\vspace{1.2cm}

{\LARGE \bf{Bumping into the species scale\vspace{0.4cm}\\ with the scalar potential}}\\

\vspace{2.2 cm} {\Large David Andriot}\\
\vspace{0.9 cm} {\small\slshape Laboratoire d’Annecy-le-Vieux de Physique Th\'eorique (LAPTh),\\
CNRS, Universit\'e Savoie Mont Blanc (USMB), UMR 5108,\\
9 Chemin de Bellevue, 74940 Annecy, France}\\
\vspace{0.5cm} {\upshape\ttfamily andriot@lapth.cnrs.fr}\\

\vspace{2.8cm}

{\bf Abstract}
\vspace{0.1cm}
\end{center}

\begin{quotation}
As a quantum gravity cut-off, the species scale $\Lambda_s$ gets naturally compared to the energy scale of a scalar potential $V$ in an EFT. In this note, we compare the species scale, its rate $|\nabla \Lambda_s|/\Lambda_s$ and their field dependence, to those of a scalar potential. To that end, we first identify a string compactification leading to a scalar potential with the same properties as the species scale, namely, being positive, starting at a maximum in the bulk of field space and going asymptotically to zero. The trajectory followed in our 14-fields scalar potential is the steepest descent. Evaluating the rate $|\nabla V|/V$ along this path, we then observe a local maximum, or bump, a feature noticed as well for the species scale. We investigate the origin of this bump for the scalar potential, and compare it to that of the species scale.
\end{quotation}

\end{titlepage}

\newpage

\tableofcontents

\section{Introduction}

The swampland program \cite{Vafa:2005ui, Palti:2019pca, vanBeest:2021lhn, Agmon:2022thq} aims at characterising $d$-dimensional theories that are effective field theories (EFT) of quantum gravity, and distinguish them from phenomenological models that admit no such UV completion. Any EFT should come with a cut-off energy scale, and for a quantum gravity EFT, that cut-off should lie below the scale at which quantum gravity effects become relevant. This scale is a priori the Planck mass $M_p$. However, in the presence of $N$ light modes, it has been argued that the quantum gravity scale is lowered to the species scale $\Lambda_s = M_p/N^\frac{1}{d-2}$. That scale was initially introduced and discussed in \cite{Dvali:2007hz, Dvali:2007wp, Dvali:2009ks, Dvali:2010vm, Dvali:2012uq}, and was revisited recently, among others, in \cite{Castellano:2021mmx, Long:2021jlv, Castellano:2022bvr, vandeHeisteeg:2022btw, Cribiori:2022nke, vandeHeisteeg:2023ubh}.

Many EFTs obtained from string theory (on a $d=4$ Minkowski spacetime) have a moduli space, i.e.~a space of massless scalar fields ${\varphi^i}$ governed by a metric $g_{ij}$. In that context, one quantum gravity effect has been introduced under the name of swampland distance conjecture \cite{Ooguri:2006in, Klaewer:2016kiy, Baume:2016psm}: when going to the asymptotics of the moduli space, a tower of massive states goes down exponentially to zero in mass. The emergent string conjecture \cite{Lee:2019wij} essentially indicates that such a tower is either populated of Kaluza--Klein states or of string states. In either case, these can be understood as quantum gravity degrees of freedom. Therefore, the typical mass scale of the tower is usually required to stand above the EFT cut-off, and can be identified with the species scale.

\begin{figure}[H]
\begin{center}
\begin{subfigure}[H]{0.45\textwidth}
\includegraphics[width=\textwidth]{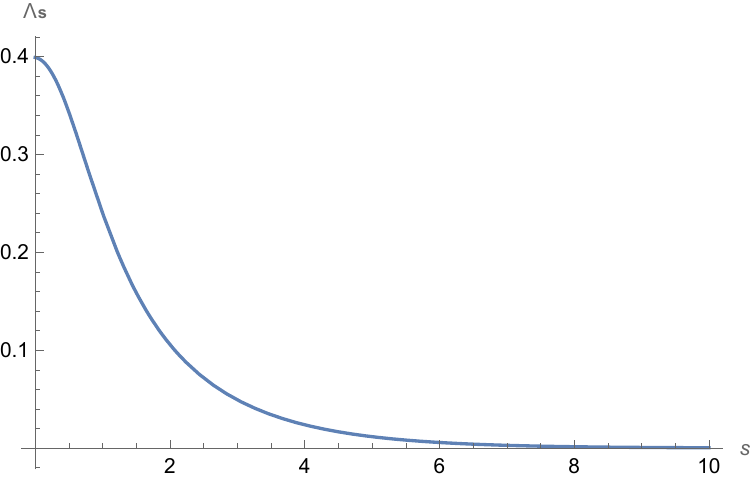}\caption{$\frac{\Lambda_s}{M_p}(\hat{s})$}\label{fig:Ls}
\end{subfigure}\qquad \quad
\begin{subfigure}[H]{0.45\textwidth}
\includegraphics[width=\textwidth]{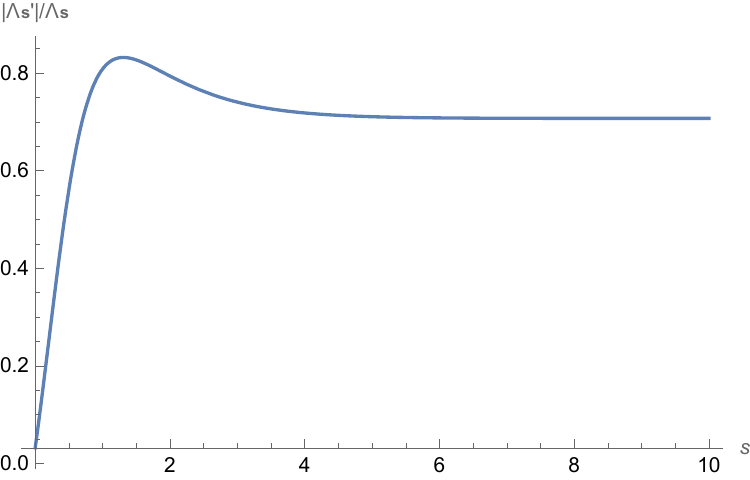}\caption{$M_p \frac{|\Lambda_s'|}{\Lambda_s}(\hat{s})$}\label{fig:Lsratio}
\end{subfigure}
\caption{Species scale $\Lambda_s$ in Planckian units as given in \eqref{Ls} in terms of the canonically normalized real field $\hat{s}$, with parameters $\alpha=2\pi$, $\beta=6$, $c =\sqrt{2}$, corresponding to example 1 of \cite{vandeHeisteeg:2023ubh}. We also give the rate $\frac{|\Lambda_s'|}{\Lambda_s}(\hat{s})$ \eqref{rateL} in Planckian units, that displays a maximum, or bump, a feature of interest in this work.}\label{fig:Ls2}
\end{center}
\end{figure}

The latter implies that the species scale can have a moduli dependence. From the first definition of the species scale, this can be understood by the fact that the number of light modes is moduli-dependent, as actually illustrated by the distance conjecture. Moreover, one deduces that $\Lambda_s\rightarrow 0$, typically exponentially, in the asymptotics; it has even been argued to start at a maximum at the ``desert point'', in the middle or bulk of the moduli space, and from there go down to zero \cite{Long:2021jlv, vandeHeisteeg:2022btw, vandeHeisteeg:2023ubh}. Last but not least, when studying this evolution, one is led to consider the variation rate $|\nabla \Lambda_s|/\Lambda_s$, where $(\nabla \Lambda)^2 \equiv g^{ij}\, \del_{\varphi^i} \Lambda\, \del_{\varphi^j} \Lambda $. It has been argued in \cite{vandeHeisteeg:2023ubh} that this rate should be bounded from above
\beq
\frac{|\nabla \Lambda_s|}{\Lambda_s} < \frac{{\cal O}(1)}{M_p} \ ,\label{rateL}
\eeq
by a number of order 1 in Planckian units. Based on type II compactifications on Calabi-Yau 3-folds (see Section \ref{sec:Ls}), it was shown that this rate does not grow monotonically to its asymptotic value, but first goes to a maximum and then diminishes. We refer to this feature as a ``bump'' in the curve, and will focus on it in this work. These properties of the species scale are summarized and illustrated in Figure \ref{fig:Ls2}.

In this work, we are interested in $d=4$ theories of scalar fields ${\varphi^i}$ minimally coupled to gravity, with a positive scalar potential $V>0$ and a field space metric $g_{ij}$
\beq
\int \d^4 x \sqrt{|g_4|} \left(\frac{M_p^2}{2} \R_4 - \frac{1}{2} g_{ij} \del_{\mu}\varphi^i \del^{\mu}\varphi^j - V \right) \ . \label{EFTV}
\eeq
Such a theory can serve as an (observationally valid) cosmological model for dark energy, and more generally for inflation or quintessence models. Constraining such theories to be quantum gravity EFTs has therefore been an active topic in the swampland program, leading to proposed characterisations of the kind
\beq
\frac{|\nabla V|}{V} \geq \frac{{\cal O}(1)}{M_p} \ ,\label{rateV}
\eeq
at least in the asymptotics of field space \cite{Obied:2018sgi, Rudelius:2019cfh, Bedroya:2019snp}. These de Sitter conjectures also agree with the general observation for $V>0$ that $V \rightarrow 0$ exponentially, in some asymptotics of field space, and \eqref{rateV} then characterises the exponential rate.

If \eqref{EFTV} is an effective theory of quantum gravity, its typical energy scales, below its cut-off, should be smaller than the quantum gravity scale. A typical energy scale of \eqref{EFTV} is that of the scalar potential, corresponding to $\sqrt{V}/M_p$, so one may bound it by a Planckian scale (see the refined TCC, and the ATCC, for a use of this argument \cite{Andriot:2022brg}). Given the above, one may be more constraining and bound it by the species scale, giving
\beq
\Lambda_s \geq \frac{\sqrt{V}}{M_p} \ . \label{LV}
\eeq
Related ideas can be found in \cite{Hebecker:2018vxz, Scalisi:2018eaz}.

Comparing the two rates \eqref{rateL} and \eqref{rateV}, one can see that these two inequalities lead (at least in field space asymptotics) to
\beq
\frac{|\nabla \Lambda_s|}{\Lambda_s} \leq \frac{|\nabla V|}{V} \ ,\label{ratecompare}
\eeq
with a possible extra ${\cal O}(1)$ factor. Importantly, \eqref{ratecompare} agrees with \eqref{LV}. Indeed, if both $\Lambda_s$ and $V$ go down to zero exponentially, and if their exponential rates are ordered as in \eqref{ratecompare}, then $\Lambda_s$ is dominant over $V$ in the asymptotics, which is in agreement with \eqref{LV}. Note that we make use here of the previously discussed properties of the species scale, but now in presence of a scalar potential, this requires to trade the moduli space for the field space; the latter is usually less well-understood, but it is generally believed that similar results should hold.

\begin{figure}[H]
\begin{center}
\begin{subfigure}[H]{0.45\textwidth}
\includegraphics[width=\textwidth]{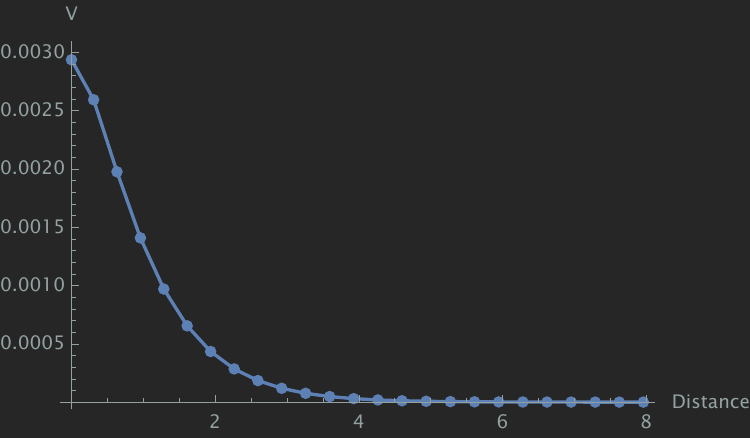}\caption{$V$}\label{fig:m55774potdist}
\end{subfigure}\qquad \quad
\begin{subfigure}[H]{0.45\textwidth}
\includegraphics[width=\textwidth]{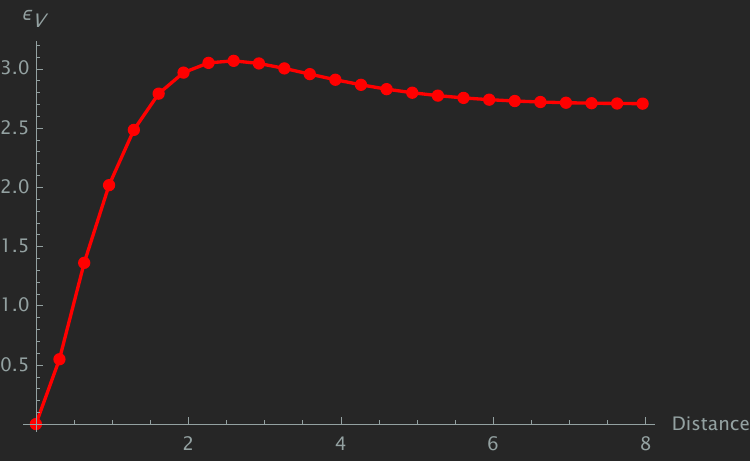}\caption{$\epsilon_V$}\label{fig:m55774epsilondist}
\end{subfigure}
\caption{Potential $V$ around the de Sitter solution $m_{5577}^+ 4$ \cite{Andriot:2022way}, in terms of the field distance, along the trajectory following the steepest descent, and parameter $\epsilon_V$ capturing the rate $|\nabla V|/V$, along the same trajectory. Interestingly, we have in \ref{fig:m55774potdist} a concrete example of a potential and field direction with the desired behaviour, and we observe in \ref{fig:m55774epsilondist} a bump in the evolution of $\epsilon_V$ along the trajectory; understanding the origin of the bump is the purpose of Section \ref{sec:bump}.}\label{fig:m55774intro}
\end{center}
\end{figure}

In this work, we are then interested in \eqref{LV} and \eqref{ratecompare}, and more generally in comparing features of $\Lambda_s$ and $V$. As recalled, $\Lambda_s$ is expected to start in field space at a maximum in the bulk and go exponentially down to zero in the asymptotics. We explain below and in Section \ref{sec:approppot} how to find an example of a scalar potential from a string compactification, having the same properties; this is already non-trivial. Then, a remaining feature is the evolution of the rate $|\nabla \Lambda_s|/\Lambda_s$, that exhibited a bump. Considering the field space trajectory given by the steepest descent of the scalar potential, we observe a similar bump in the potential rate $|\nabla V|/V$, or equivalently in $\epsilon_V = 1/2\, (M_p\, |\nabla V|/V)^2$. This is displayed in Figure \ref{fig:m55774intro}. In Section \ref{sec:bump}, we then aim at understanding the origin of this bump for $V$, and whether it can be related to the one for $\Lambda_s$.

To be able to compare features of $\Lambda_s$ and $V$, a starting point is to have a positive scalar potential, $V>0$, ideally with a maximum in the bulk of field space, i.e.~a de Sitter solution. Obtaining a trustable scalar potential from a string theory compactification, that is positive, is already challenging; having in addition a de Sitter critical point is notably difficult \cite{Bena:2023sks}, in line with the de Sitter swampland conjectures. A first example is non-supersymmetric GKP compactifications \cite{Giddings:2001yu}, that give rise to a positive scalar potential (see e.g.~\cite{Calderon-Infante:2022nxb} for a recent extensive study). Those are however known not to admit de Sitter solutions, without extra ingredients (see e.g.~\cite[(6.18)]{Andriot:2017jhf} for a related no-go theorem). Another example are compactifications on group manifolds: these consistent truncations of 10d type II supergravities provide examples of de Sitter solutions (see \cite{Andriot:2022way} for a recent classification), and thus of positive scalar potentials around each solution. We consider here an example of such a type IIB compactification, solution $m_{5577}^+ 4$ found and studied in \cite{Andriot:2022way, Andriot:2022yyj, Andriot:2022bnb}. It has $O_5/D_5$ and $O_7/D_7$, and gives rise to a 4d theory \eqref{EFTV} with 14 scalar fields: 7 axions and 7 saxions, the latter corresponding to the 6 radii and the dilaton fluctuations.

In agreement with de Sitter conjectures \eqref{rateV}, forbidding de Sitter solutions in the asymptotics of field space, these examples of de Sitter critical points are believed to stand in the bulk. As 10d supergravity solutions, they should however be in the classical regime (large volumes, small string coupling, etc.) to be valid string backgrounds, and this is in tension with them being in the bulk of field space \cite{Banlaki:2018ayh, Junghans:2018gdb, Andriot:2019wrs, Grimm:2019ixq, Andriot:2020vlg}. They actually seem to lie at the edge of classical validity, whenever this can be checked \cite{Andriot:2020vlg}; the lattice conditions, related to the compactness requirements of the group manifolds, actually make it difficult to verify these requirements in practice. We refrain from studying this aspect for the solution of interest, $m_{5577}^+ 4$: we assume that it is enough in the bulk to mimic the behaviour of $\Lambda_s$, but that this solution and its scalar potential are still mostly in a trustable regime. Note also that the field trajectory to be followed is in the direction of the classical regime.

A second point is to have a field direction that starts at a maximum and asymptotes to zero, as in Figure \ref{fig:Ls} for $\Lambda_s$ and \ref{fig:m55774potdist} for $V$. This is way less trivial than it seems in multifield potentials. Indeed, in most compactifications that give rise to a potential, some fields are stabilized, while others remain flat, none of which matches the desired behaviour. The dilaton potential is known to asymptote to zero \cite{Dine:1985he}, but the dilaton is typically stabilised at a minimum in a de Sitter solution; we will make this manifest in Figure \ref{fig:3dpotphi} and \ref{fig:3drhotau}. Here, we will then be guided by the de Sitter tachyonic direction \cite{Danielsson:2012et, Andriot:2021rdy}, and will comment more in Section \ref{sec:m55774} on its asymptotics.

Finally, in a multifield scalar potential, one should pick a field trajectory along which to study the evolution of the rate $|\nabla V|/V$ or $\epsilon_V$. We choose to consider the steepest descent, defined by following the path that minimizes the most the potential; we do so numerically. Note that this trajectory starts along the tachyonic direction mentioned above, but then typically deviates from it. While the steepest descent trajectory can be defined universally, it also has the advantage to get the tightest constraints on the potential, since it minimizes it. This is interesting in view of \eqref{LV}, as the species scale could then also get minimized and thus bring a tight cut-off scale. The result is that along this trajectory, we observe a bump in $\epsilon_V$: see Figure \ref{fig:m55774epsilondist}, similarly to the one for $\Lambda_s$ in Figure \ref{fig:Lsratio}. The steepest descent trajectory is however difficult to handle analytically (see Section \ref{sec:steep}), and it is then the purpose of Section \ref{sec:bump} to understand how a bump can be generated along that path.\\

This note is organised as follows. We first provide more details on the species scale and its moduli dependence in Section \ref{sec:Ls}. We then explain in Section \ref{sec:approppot} how to find a de Sitter solution, its scalar potential and a field direction that exhibit the desired properties. We end up picking the de Sitter solution $m_{5577}^+ 4$ that we present, together with its scalar potential. We further discuss the steepest descent procedure and its results in Section \ref{sec:steep}. We then turn to finding the origin of the bump in $\epsilon_V$, displayed in Figure \ref{fig:m55774epsilondist}, in Section \ref{sec:bump}. To that end, we investigate the dependence of the potential (exponential or linear) on the various fields that seem to contribute to the steepest descent trajectory. Two lemmas presented in Section \ref{sec:expan}, on properties of scalar potentials with pure exponential dependence, are proven in Appendix \ref{ap:proofs}. We conclude with an outlook in Section \ref{sec:outlook}.

\section{Setting the stage}

\subsection{The species scale}\label{sec:Ls}

Associating the species scale to the typical mass of a quantum gravity tower, and using the distance conjecture, one gets a modulus-dependent expression for the species scale, valid in the asymptotics: essentially a decaying exponential. In \cite{vandeHeisteeg:2022btw, Cribiori:2022nke, vandeHeisteeg:2023ubh}, the species scale was argued to appear in higher derivative corrections of a quantum gravity EFT action. In a specific class of string theory compactification (type II on Calabi-Yau 3-fold), leading to an ${\cal N}=2$ 4d theory, one such higher derivative correction was analysed, and a modulus-dependent expression could then be found for the species scale. Namely, it was proposed that $\Lambda_s= \frac{M_p}{\sqrt{F_1}}$, where $F_1$ is the one-loop topological free energy; we ignore for the present discussion a possible extra constant and we refer to \cite{vandeHeisteeg:2022btw, vandeHeisteeg:2023ubh} for more details. The properties of $F_1$ allowed to get an expression admitting valid corrections beyond the asymptotic exponential. Indeed, a general expression for the species bound was then proposed
\beq
\Lambda_s= \frac{M_p}{\sqrt{\alpha}}\ \frac{1}{\sqrt{e^{c \hat{s}} - \frac{\beta c}{\alpha} \hat{s} }} \ , \label{Ls}
\eeq
where matching with \cite[(4.33)]{vandeHeisteeg:2023ubh} gives $\alpha= \frac{2\pi}{12}c_2$ and $s=e^{c \hat{s}}$. The real field $\hat{s}$ is canonically normalized and $c>0$ depends on the example considered; furthermore, $\alpha>0, \beta>0$ in most examples. The above expression is given in the limit of large field $\hat{s} \rightarrow \infty$, but nothing qualitatively changes when considering $\Lambda_s$ for small $\hat{s}$ values (see the graphs in \cite{vandeHeisteeg:2023ubh}), so we consider for now that expression in full generality.

Expanding the above at large $\hat{s}$, one gets
\beq
\Lambda_s \approx \frac{M_p}{\sqrt{\alpha}} \left( e^{-\frac{c}{2} \hat{s}} +\frac{1}{2} \frac{\beta c}{\alpha} e^{-\frac{3c}{2} \hat{s}} \ \hat{s} \right)\ . \label{Lsapprox}
\eeq
Importantly, note that this expansion is also valid at small $\hat{s}$ values. We see that beyond the asymptotically leading exponential, the subleading one is coupled to a linear field dependence. In other words, we have a deviation from a pure exponential expression. This deviation (corresponding to a {\it log} correction in \cite{vandeHeisteeg:2023ubh}), and especially its positive sign, are responsible for the bump, as discussed in \cite{vandeHeisteeg:2023ubh}.

Indeed, one gets from \eqref{Lsapprox} in the large field limit $\hat{s} \rightarrow \infty$
\beq
-\frac{\Lambda_s'}{\Lambda_s} \approx \frac{c}{2} \left( 1 + \frac{\beta c}{\alpha} e^{-c \hat{s}} \ \hat{s} \right)\ .
\eeq
This ratio then approaches its asymptotic value $\frac{c}{2}$ from above since the second term is {\sl positive}. In addition, using that \eqref{Lsapprox} is also valid for small $\hat{s}$ values, one verifies that the ratio equates $\frac{c}{2} \left( 1 - \frac{\beta }{\alpha}\right)$ at $\hat{s}=0$, i.e.~less than the asymptotic value. One concludes that there must be a maximum in between, corresponding to the bump. This is illustrated in Figure \ref{fig:Lsratio}.

In the remainder of this work, we investigate whether the bump observed in $\epsilon_V$, with the scalar potential of solution $m_{5577}^+ 4$, can be explained in the same way.

\subsection{The scalar potential and its steepest descent}\label{sec:m55774}

\subsubsection{Finding an appropriate scalar potential}\label{sec:approppot}

As motivated in the Introduction, we are interested in positive scalar potentials obtained from string theory compactifications, that exhibit a de Sitter critical point. In addition, we want a field direction that starts there at a maximum (therefore as a tachyon) and goes to zero asymptotically. To find such a setting, we start with the database of de Sitter solutions found in 10d type IIA and IIB supergravities with $D_p$-branes and orientifold $O_p$-planes on group manifolds \cite{Andriot:2020wpp, Andriot:2021rdy, Andriot:2022way}, and classified in \cite{Andriot:2022way}. We first restrict to only those obtained on a {\sl compact} group manifold, as determined in \cite{Andriot:2022yyj}. We then consider the whole set of scalar fields and 4d theory obtained by a consistent truncation in \cite{Andriot:2022bnb}; note that this information, in particular the corresponding scalar potential, is derived automatically thanks to the code {\tt MSSV}. A 10d de Sitter solution corresponds in that setting to a positive critical point of the 4d scalar potential. As is common \cite{Andriot:2021rdy}, all these solutions have 4d tachyonic directions; for our purposes, we further restrict to the solutions which only have {\sl one} such unstable direction, as determined in \cite[Tab. 5,6,7]{Andriot:2022bnb}. We are left with the following solutions
\beq
s_{55}^+ 14-17, 20,21,\ s_{66}^+ 1,\ s_{6666}^+ 1,3,4,\ m_{46}^+ 10,\ m_{5577}^+ 3-6, \ m_{5577}^{*+} 1 \ . \label{sol0}
\eeq

To find an example that serves our purposes, we are left to look at the asymptotics, and verify that it goes to zero. This requires to choose a field direction: the simplest here is to consider again the (single) tachyonic direction, even though we will later deviate from it with the steepest descent. Let us explain how we study the profile of the potential along the tachyonic direction, following \cite[Sec. 2.2]{Andriot:2022bnb} and \cite{Andriot:2021rdy}. The tachyon is given as an eigenvector of the mass matrix, expressed in an orthonormal canonical basis. We denote it by $\del_{\hat{t}}$, and the orthonormal eigenvectors by $\del_{\hat{t}^k_{\bot}}$. We associate to them the fields $\hat{t}, \hat{t}^k_{\bot}$ and fix their origins as the values $\hat{t}_0, \hat{t}^k_{\bot 0}$ at the de Sitter critical point; the $\hat{t}^k_{\bot}$ are by definition stabilized (or massless) at the critical point. We consider the following potential, developed as a Taylor series along the tachyonic direction
\beq
V(\hat{t}_0 + y , \hat{t}^k_{\bot 0}) = V_0 + \sum_{n>0} \frac{y^n}{n!} [\del_{\hat{t}}^n V ]_0 \ ,\qquad y = \hat{t} - \hat{t}_0 \ .
\eeq
In addition, the tachyonic eigenvector is expressed in the standard field basis via the following equalities
\beq
\del_{\hat{t}} = v^i \del_{\hat{\varphi}^i} = (P^{-1})^j{}_i v^i \del_{\varphi^j} \equiv w^i  \del_{\varphi^i} \ ,
\eeq
where $\varphi^i$ are the standard scalar fields and $\hat{\varphi}^i$ their counterparts in a canonical basis (where $g_{ij}=\delta_{ij}$). Note that the $v^i$ are normalised to 1, but not the $w^i$. We thus rewrite the above expansion as
\bea
V(\hat{t}_0 + y , \hat{t}^k_{\bot 0}) & = V_0 + \sum_{n>0} \frac{y^n}{n!} [( w^j  \del_{\varphi^j})^n V(\varphi^i) ]_0 \nn \\
& = V_0 + \sum_{n>0} \frac{y^n}{n!} [\del_y^n V(\varphi^i_0 +\, w^i y) ]_0 = V(\varphi^i_0 + w^i\, y) \ , \label{Vtac}
\eea
where we used that $\del_{\hat{t}} = \del_y$ and $\del_y =\del_y \varphi^i \, \del_{\varphi^i}$. This shows that $V(\varphi^i_0 + w^i\, y)$ corresponds to the potential along the tachyonic direction.

\begin{figure}[H]
\begin{center}
\begin{subfigure}[H]{0.45\textwidth}
\includegraphics[width=\textwidth]{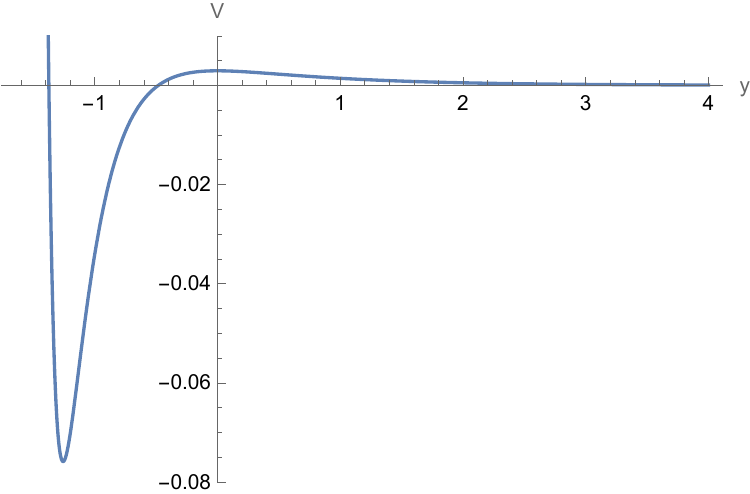}\caption{$m_{5577}^+ 4$}\label{fig:Vm55774}
\end{subfigure}\qquad \quad
\begin{subfigure}[H]{0.45\textwidth}
\includegraphics[width=\textwidth]{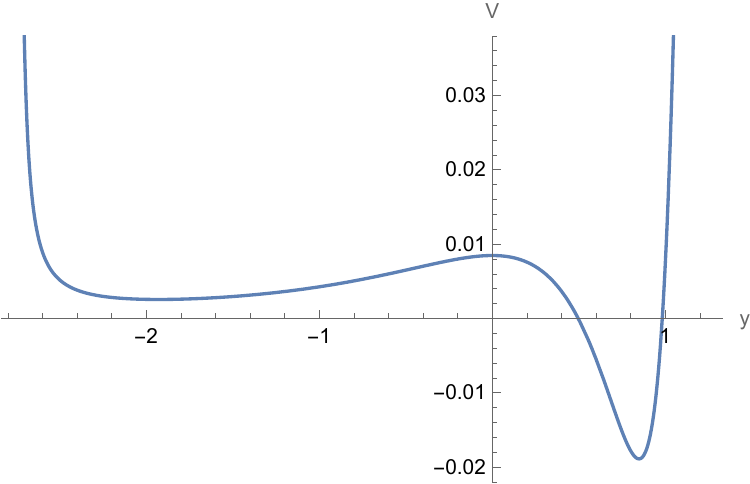}\caption{$s_{6666}^+ 4$}\label{fig:Vs66664}
\end{subfigure}
\caption{Potential $V$ along the tachyonic direction $y$ for two de Sitter solutions (at $y=0$). Only the potential of solution $m_{5577}^+ 4$ exhibits a positive runaway tachyonic direction.}\label{fig:Vtac}
\end{center}
\end{figure}

We then display the potential along the single tachyonic direction for the solutions \eqref{sol0}, as in Figure \ref{fig:Vtac}. This is made possible thanks to the code {\tt MSSV} \cite{Andriot:2022bnb} which identifies the tachyonic direction, and can compute the potential along it using \eqref{Vtac}. To our surprise, many solutions do not have runaway tachyonic directions in the asymptotics, but end up being divergent. Only some type IIB solutions allow for a tachyonic direction that goes down to zero positively and asymptotically. Those are
\beq
s_{55}^+ 14-17,\ m_{5577}^+ 3-6, \ m_{5577}^{*+} 1 \ .\label{sol1}
\eeq
It would be interesting to understand better this difference with the other solutions. We then focus on one of the solutions \eqref{sol1}, namely $m_{5577}^+ 4$.

Solution $m_{5577}^+ 4$ is a type IIB de Sitter solution on a 6d compact group manifold. The latter is based on the solvable Lie algebra $\mathfrak{g}_{6.92}^{0, \mu_0,\nu_0}$. The solution also admits non-zero $F_1, F_3$ and $H$ fluxes. Finally, it has 2 intersecting sets of $O_5/D_5$ and 2 of $O_7/D_7$. We refer to \cite{Andriot:2022way, Andriot:2022yyj} for more details; in particular, the whole solution (including flux components) can be found in \cite[App. C.1]{Andriot:2022way}.

Thanks to the consistent truncation of \cite{Andriot:2022bnb} and the code {\tt MSSV}, we obtain the corresponding 4d theory (field space metric, scalar potential) for the 14 real scalar fields. Those can be understood as fluctuations of the 10d fields, with the following components
\begin{equation}
\begin{aligned}
       m_{5577}: \quad {\rm saxions:}\quad &g_{11} \,, g_{22} \,, g_{33} \,, g_{44} \,, g_{55}  \,, g_{66}\,, \phi \,, \\
       {\rm axions:}\quad &C_{2 \ 12} \,, C_{2 \ 34}\,, C_{4 \ 1356}\,, C_{4 \ 2456} \,, b_{14}\,, b_{23}\,, g_{56}\,.
\end{aligned}\label{fields}
\end{equation}
We can then compute the parameters $\epsilon_V$ and $\eta_V$ at any given point in field space, and we will do so along the trajectory. We recall the definitions
\beq
\epsilon_V = \frac{M_p^2}{2} \left( \frac{\sqrt{g^{ij} \partial_i V  \partial_j V }}{V} \right)^2 \ ,\ \eta_V = M_p^2\ \frac{\text{min} \left( \, g^{ij} \nabla_j \partial_k V \, \right)}{V}\,, \label{par}
\eeq
where $\nabla_j v_k = \del_j v_k - {\Gamma^l}_{jk} v_l$ is the covariant derivative on $v_k$ and
${\Gamma^l}_{jk}$ denotes the Christoffel symbol associated with $g_{ij}$. The mass matrix has as a component $M^i{}_k = g^{ij} \nabla_j \partial_k V$ and ``min'' denotes its minimal eigenvalue, its eigenvalues being the masses${}^2$. By definition, $\epsilon_V$ is vanishing at the critical point, and can only grow when going away from it. Regarding $\eta_V$, we compute it at the critical point to be $\eta_V = -3.7733$ \cite{Andriot:2022bnb}. Its evolution away from the critical point is less obvious, and we will display it.

\subsubsection{Steepest descent procedure and results}\label{sec:steep}

\begin{figure}[H]
\begin{center}
\begin{subfigure}[H]{0.45\textwidth}
\includegraphics[width=\textwidth]{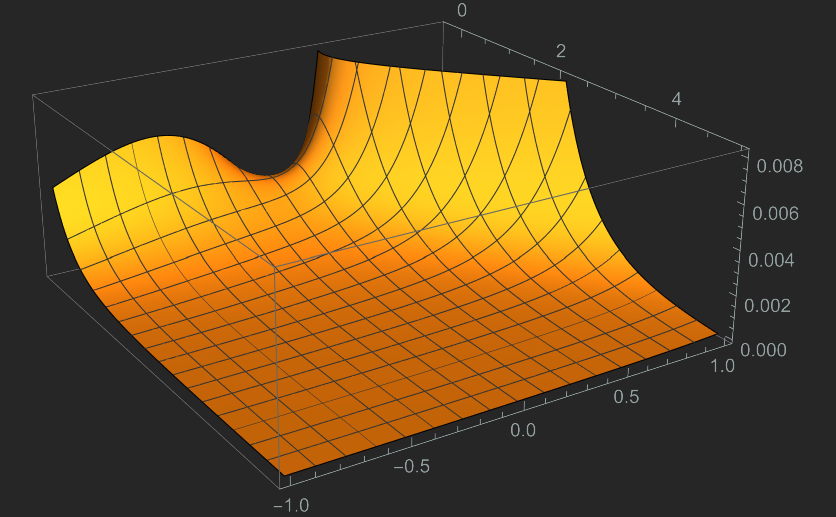}\caption{$V(y,\phi)$}\label{fig:3dpotphi}
\end{subfigure}\qquad \quad
\begin{subfigure}[H]{0.45\textwidth}
\includegraphics[width=\textwidth]{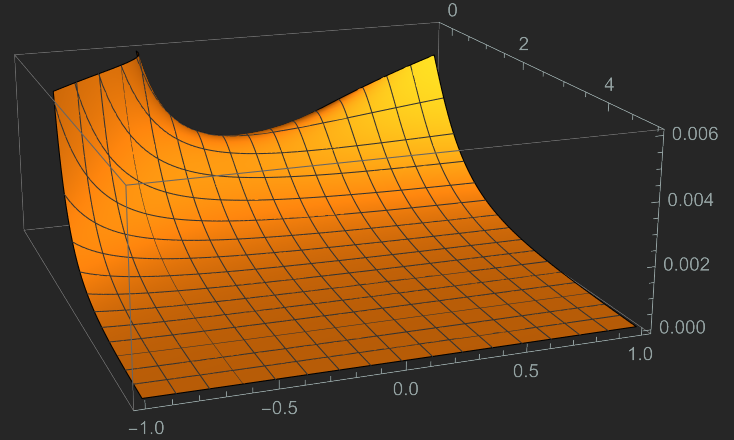}\caption{$V(y,g_{11})$}\label{fig:3dpotg11}
\end{subfigure}
\caption{Scalar potential around the solution $m_{5577}^+ 4$ (corresponding to the critical point), displayed along the tachyonic direction $y$, and $\phi$ or $g_{11}$.}\label{fig:3dpot}
\end{center}
\end{figure}

We now make use of the 4d theory described previously (scalar potential, field space metric), based on the de Sitter solution $m_{5577}^+ 4$. Let us first display the scalar potential along some directions in Figure \ref{fig:3dpot}.
We now determine the trajectory going down the potential from the critical point and following the steepest descent. We hope this way to asymptote to zero, although there is a priori no guarantee to get the same asymptotics as the tachyonic direction. To follow this path, we create a step by step algorithm, finding at each step the point minimizing the potential in a given neighborhood. We build this way, point by point, a trajectory in field space corresponding to the steepest descent. In the end, the algorithm works, in the sense that the potential does diminish to zero, and we can compute the potential, $\epsilon_V$ and $\eta_V$ at each point on the trajectory. We make it run on 24 points, whose initial (critical point) and final values are
\bea
&V_0 = 0.0029345\ , \ \epsilon_V=0 \ ,\ \eta_V= -3.7733 \ , \label{fields0}\\
&g_{11}=1 \,, g_{22}=1 \,, g_{33}=1 \,, g_{44}=1 \,, g_{55}=1  \,, g_{66}=1\,, \phi=0 \,, \nn\\
&C_{2 \ 12}=0 \,, C_{2 \ 34}=0\,, C_{4 \ 1356}=0\,, C_{4 \ 2456}=0 \,, b_{14}=0\,, b_{23}=0\,, g_{56}=0\,,\nn\\
&\nn\\
&V_{24} = 8.5940 \cdot 10^{-8}\ , \ \epsilon_V=2.7051 \ ,\ \eta_V= -0.010039 \ ,  \label{fields24}\\
&g_{11}=2.0857 \,, g_{22}=1.7979 \,, g_{33}=1.9613 \,, g_{44}=2.2293 \,, g_{55}=2.8996  \,, g_{66}=1.961\,, \phi=-4.4618 \,, \nn\\
&C_{2 \ 12}= -0.050124 \,, C_{2 \ 34}=0.034256\,, C_{4 \ 1356}=0.0015444\,, C_{4 \ 2456}=0.0015444 \,,\nn\\
&b_{14}=-0.39297\,, b_{23}=0.61030\,, g_{56}=-0.017484\,.\nn
\eea
By definition, the steepest descent first follows the tachyonic direction, but eventually moves away from it.

Mathematically, going along the steepest descent of the potential corresponds to a trajectory along the (opposite) gradient direction, $-\overrightarrow{\nabla} V$: see e.g.~\cite{Lambers} for a proof. This means that field directions orthogonal to the gradient (meaning along which the gradient is vanishing) are also orthogonal to the trajectory. In other words, the gradient orthogonal to the trajectory direction has to vanish. In an earlier version of this work, we attempted at evaluating the gradient along the trajectory and orthogonal to it, but made a mistake due to a wrong use of the command {\tt Orthogonalize}. For this version, we could verify numerically that the trajectory direction is indeed very close to that of the gradient. Note also that having transverse vanishing gradients does not mean we are in a valley: the steepest descent can also be followed on a concave surface, like a sphere. This is also illustrated by the computation of $\eta_V$ along the trajectory.

\begin{figure}[h]
\begin{center}
\begin{subfigure}[H]{0.45\textwidth}
\includegraphics[width=\textwidth]{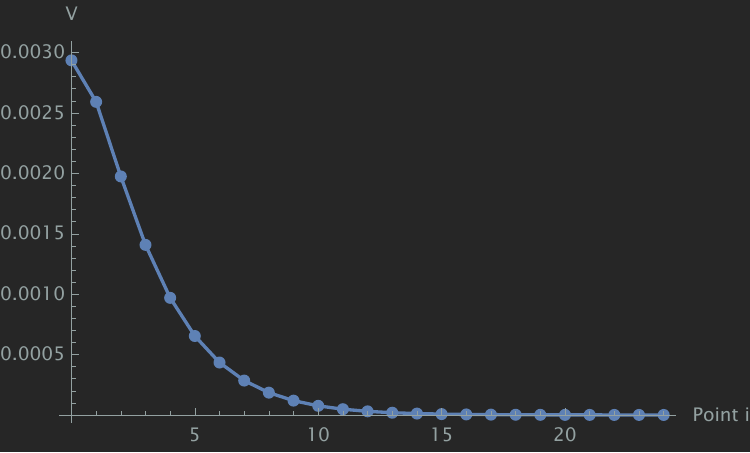}\caption{$V$}\label{fig:m5577pot}
\end{subfigure}\qquad \quad
\begin{subfigure}[H]{0.45\textwidth}
\includegraphics[width=\textwidth]{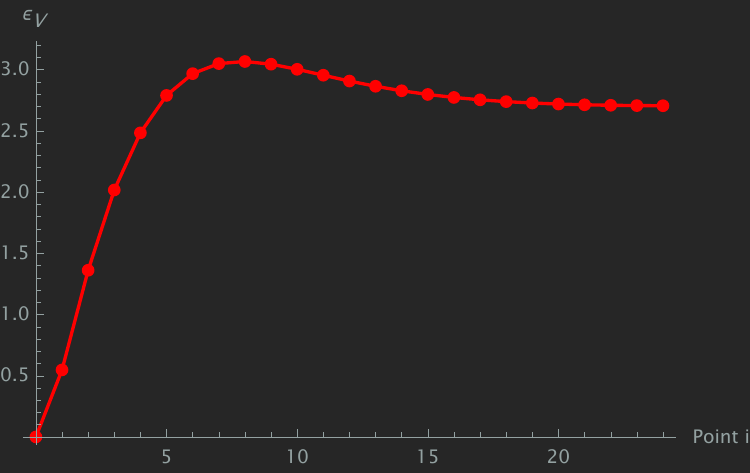}\caption{$\epsilon_V$}\label{fig:m5577epsilon}
\end{subfigure}\\
\begin{subfigure}[H]{0.45\textwidth}
\includegraphics[width=\textwidth]{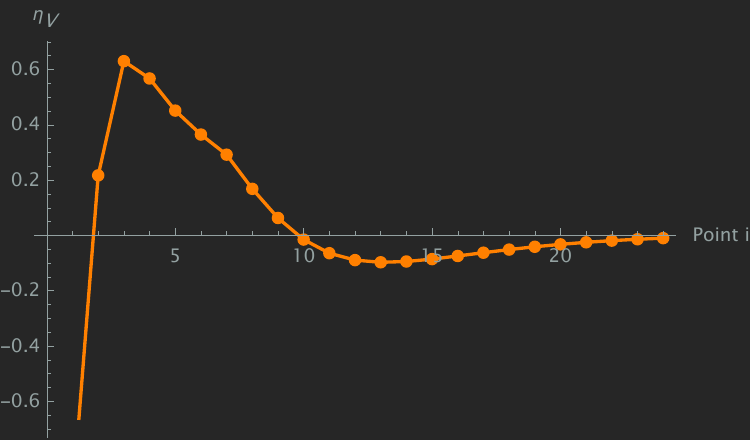}\caption{$\eta_V$}\label{fig:m5577eta}
\end{subfigure}
\caption{Results of the steepest descent for the scalar potential around the de Sitter solution  $m_{5577}^+ 4$.}\label{fig:m5577}
\end{center}
\end{figure}

The results of this steepest descent procedure are given in Figure \ref{fig:m5577}. All details can be found in the corresponding notebook {\tt Steepest descent full m5577 4 MSSV}, based on the code {\tt MSSV}.
\noindent As discussed already, the main result is the bump appearing in $\epsilon_V$ in Figure \ref{fig:m5577epsilon}. It will be the focus of Section \ref{sec:bump} to understand its origin.\\

We end this section by explaining a few technicalities on the computation of the field distance, and with more comments on the steepest descent. Note that the graphs in Figure \ref{fig:m55774intro} and \ref{fig:m5577}, given in terms of the field distance or in terms of the points sample, show almost no difference. In the following, we then only display graphs in terms of the points sample, which is much simpler.

Two successive points along the trajectory, $\varphi [k]$ and $\varphi [k+1]$, given by the coordinates or field values $(\varphi^i)$, define a ``vector'' along the trajectory by taking $\varphi [k+1] - \varphi [k]$. Because of this difference, this ``vector'' rather corresponds mathematically to a one-form on the basis ${\d \varphi^i}$, having in mind that the discretization steps are small, we identify this ``vector'' as this one-form in the following. Denoting the trajectory direction $s$ and the displacement $\d s$, one has
\beq
\d \varphi^i = \frac{\del \varphi^i}{\del s} \d s \ .
\eeq
The above quantities appear also as the components of the vector along the trajectory
\beq
\del_s = \frac{\del \varphi^i}{\del s} \del_{\varphi^i} \ .
\eeq
The field space distance along the trajectory is thus
\beq
g_{ij} \d \varphi^i \d \varphi^j = g_{ij} \frac{\del \varphi^i}{\del s} \d s \frac{\del \varphi^j}{\del s} \d s \equiv g_{ss} {\d s}^2 \ .\label{dist}
\eeq
Therefore, if we consider $s$ to be canonically normalized, one has $g_{ss} = 1$ and \eqref{dist} provides precisely the normalized distance $\d s$: to get the latter we simply have to get the norm of the ``vector'' $\varphi [k+1] - \varphi [k]$ with the field space metric, given the left-hand side of \eqref{dist}.

We now comment more on the steepest descent trajectory. As explained above, it corresponds to following the (opposite) gradient of the potential $-\overrightarrow{\nabla} V$, i.e.~having a vanishing gradient in directions orthogonal to the trajectory. As shown in \cite[Sec. 2.5]{Calderon-Infante:2022nxb} (see also \cite{Achucarro:2018vey}), this is actually a geodesic trajectory. In \cite[Sec. 2.5]{Calderon-Infante:2022nxb}, it is also shown that this is equivalent to having a gradient flow. The latter refers to having the trajectory direction, given by $\del_{s}{\varphi}^i$ with a parameter $s$, being parallel (i.e.~proportional) to the gradient.

In this discussion, a subtlety is the notion of orthogonality, which requires the field space metric, a data independent of $V$. The gradient $\overrightarrow{\nabla} V$ should be understood as a vector of $i$-component $g^{ij} \del_j V$. With this upper index, it can be proportional to $\del_{s}{\varphi}^i$, giving a gradient flow. Considering a field direction $t$, the vectors inner product, testing e.g.~the orthogonality or the parallelism to the gradient, is also expressed via the metric, with $\overrightarrow{\del_t \varphi} \cdot  \overrightarrow{\nabla} V  = \del_t {\varphi}^i\, g_{ij}\, g^{jk} \del_k V = \del_t {\varphi}^i\, \del_i V = \del_t V $. The proof of \cite{Lambers} can also be adapted using this definition for $\overrightarrow{\nabla} V$ and this inner product.

In \cite{Calderon-Infante:2022nxb}, it is shown via the equations of motion that the slow-roll regime leads to a gradient flow trajectory. This can be understood physically: if the field rolls down very slowly (e.g.~due to a very high friction), it will go along the steepest descent path, consistently with the above. This is however unlikely to happen initially here, when leaving our de Sitter critical point: indeed, the slope quickly becomes steep while the friction diminishes, so the field first catches up speed; the slow-roll approximation is thus easily violated. Then, the field moves fast in one direction, non-geodesically, and leaves the steepest descent path. In the asymptotics though, a slow-roll regime is likely to get restored, due to the cumulated friction. It is also the asymptotics that are considered in \cite{Calderon-Infante:2022nxb}. We conclude that the steepest descent path followed here is unlikely to be a physical trajectory; it remains an interesting path to study the properties of the potential.

\section{Understanding the bump}\label{sec:bump}

In this section, we aim at understanding for the scalar potential the origin of the maximum in the curve of $\epsilon_V$ along the trajectory (see Figure \ref{fig:m55774epsilondist}), in other words, the bump. We start in Section \ref{sec:expan} by considering and analysing a potential given by a sum of exponentials. We apply this in Section \ref{sec:2fields} to a two fields, simplified model, for the solution $m_{5577}^+ 4$ and its potential. We then turn in Section \ref{sec:axion} to a linear contribution to the potential due to the axion. We finally consider in Section \ref{sec:moreexp} a more complicated model for the $m_{5577}^+ 4$ and its potential, built again only of exponentials. The latter will provide the origin of the bump.

The trajectory considered starts from the de Sitter critical point, corresponding to solution $m_{5577}^+ 4$, and evolves in the corresponding multifield scalar potential $V$ following the steepest descent. By definition, since $m_{5577}^+ 4$ admits only one tachyonic direction, the trajectory is initially along the tachyon. It is well-known that de Sitter tachyons are typically along the dilaton and volumes, denoted $(\tau,\rho,\sigma_I)$ \cite{Danielsson:2012et}, and this was verified to be the case for that solution \cite{Andriot:2022yyj, Andriot:2022bnb}. These fields can also be viewed as saxions, i.e.~radii and dilaton. This is a first reason to think that the steepest descent trajectory, starting along the tachyon, will take most of its contributions from those fields. In addition, we can verify this explicitly on the trajectory \eqref{fields24}: most of the contributions to the field direction are indeed along $g_{aa}$ and $\phi$, with a few more from the $B$-field, a point we will come back to in Section \ref{sec:axion}. We keep this intuition on the trajectory in mind, when trying to find an origin to the bump.

\subsection{Exponentials}

As just explained, most of the trajectory is along saxions. In terms of canonically normalized fields, they appear in the scalar potential through exponentials. In this subsection, we then restrict ourselves to a potential given only as a sum of exponentials.

\subsubsection{General analysis}\label{sec:expan}

We consider $V$ to be expressed in terms of canonically normalized fields $\{ \hat{\varphi}^i \}$. For simplicity, we denote the trajectory field direction as one of them, $\hat{\varphi}$. The potential of interest can then be expressed as follows
\beq
V(\hat{\varphi}) = \sum_{i=1}^n A_i\ e^{a_i \hat{\varphi}} \ ,\ a_1 < \dots < a_n < 0 \ ,\ A_n >0 \ , \label{potsingle}
\eeq
with $n\geq 2$; $n=1$ is a decreasing exponential that does not allow for a de Sitter critical point. We focus on the case where $V(\hat{\varphi}) \rightarrow 0$ for $\hat{\varphi} \rightarrow \infty$, this is why all rates $a_i$ are negative, and we then order the potential terms accordingly as in \eqref{potsingle}. We also want $V(\hat{\varphi}) \rightarrow 0$ while remaining positive, hence the sign of the asymptotic dominant term $A_n >0$. Finally we consider the last de Sitter extremum before asymptotics (necessarily a maximum) to be located at $\hat{\varphi}=0$; any other value can be obtained by redefining the constants $A_i$. Therefore, after that point, $V'$ does not change sign: it is negative as in the asymptotics, the potential decreases from the de Sitter point to zero in the asymptotics. In short, the potential \eqref{potsingle} can be a model for the complete potential; it reproduces the behaviour of Figure \ref{fig:m55774potdist}.

Note that the coefficients $A_i$ can be taken to be non-constant and depend on the other (canonically normalized) fields. It does not alter the reasoning below. In particular, the second derivative in $\hat{\varphi}$ does not need to involve a covariant derivative, because the Christoffel symbols vanish in the canonical basis.

With such a scalar potential, we are now interested in evolution of  $\epsilon_V = \tfrac{M_p^2}{2} |\nabla V|^2/V^2 $ along the trajectory. Following the steepest path amounts to follow the largest (in absolute value) gradient direction of $V$. As a consequence, the quantity $|\nabla V|^2 = \ssum \delta^{ij} \del_{\hat{\varphi}^i} V \del_{\hat{\varphi}^j} V $ admits one dominant term, ${V'}^2$, where $V' = \del_{\varphi} V$. As a first approximation, we could consider that term alone. The evolution along the trajectory of  $\epsilon_V$ then amounts to that of $|V'|/V$, on which we now focus. The problem becomes essentially single field, even though the potential is not.

By definition, the ratio $|V'|/V = -V'/V$ vanishes at $\hat{\varphi}=0$, grows then positive and asymptotes at the value $|a_n|$. Whether or not it admits a maximum in between, i.e.~a bump requires its first derivative to vanish. One has
\beq
\left(-\frac{V'}{V} \right)' = \frac{{V'}^2 - V\, V''}{V^2} \ .
\eeq
Defining $X_i= A_i\ e^{a_i \hat{\varphi}}$, one gets
\bea
{V'}^2 - V\, V'' & = (\sum_i a_i X_i)^2 - (\sum_i X_i) (\sum_j a_j^2 X_j) \label{bla}\\
&= \sum_{i \neq j} (a_i a_j - a_j^2) X_i X_j = -\sum_{i< j} (a_i - a_j)^2 X_i X_j \ . \nn
\eea

At the de Sitter critical point, one gets $\ssum a_i A_i =0$. Since $A_n >0$, this implies that there must be an $i\neq n$ such that $A_i<0$. Let us first focus on the case $n=2$: we deduce that $A_1<0$. Then we get that $X_1 X_2 <0$, and that ${V'}^2 - V\, V'' = - (a_1 - a_2)^2 X_1 X_2 > 0$. Thus
\beq
n=2:\quad \left(-\frac{V'}{V} \right)' > 0 \ .
\eeq
The ratio of interest $|V'|/V $ just grows from 0 to its asymptotic value, and there is no bump. The bump can then only appear for $n \geq 3$.

\begin{figure}[H]
\begin{center}
\includegraphics[width=0.75\textwidth]{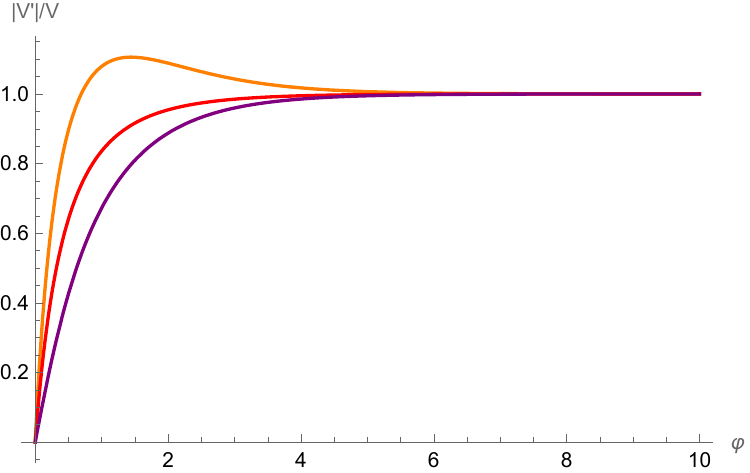}
\caption{Ratio $\frac{|V'|}{V}$ in terms of $\hat{\varphi}$ for the exponential potential \eqref{potsingle} and $n=3$, with $a_1=-3,\, a_2=-2,\, a_3=-1$ and $V_0 = 1$. Thanks to \eqref{Aiexp}, fixing $V_0''$ then fixes the coefficients. The orange curve stands for $V''_0=-4$ giving $A_1=-1, A_2= 1, A_3= 1$; the red one for $V''_0=-2.5$, $A_1=-0.25, A_2= -0.5, A_3= 1.75$, and the purple one for $V''_0=-1$, $A_1=0.5, A_2= -2, A_3= 2.5$. As proven, only the first case provides a bump.}\label{fig:n=3}
\end{center}
\end{figure}

For $n=3$, using equations at $\hat{\varphi}=0$, one can express the coefficients $A_{1,2,3}$ in terms of the corresponding values of the potential and its second derivative, $V_0>0$ and ${V''}_0<0$, as follows
\beq
A_1 = \frac{a_2 a_3 V_0 + {V''}_0}{(a_2-a_1)(a_3-a_1)} \ ,\ A_2 = \frac{-(a_1 a_3 V_0 + {V''}_0)}{(a_2-a_1)(a_3-a_2)} \ ,\ A_3 = \frac{a_1 a_2 V_0 + {V''}_0}{(a_3-a_2)(a_3-a_1)}  \ . \label{Aiexp}
\eeq
To verify whether or not we get a bump, we should check whether \eqref{bla} vanishes; note that at $\hat{\varphi}=0$, this quantity is equal to $-V_0\, {V''}_0$, which is positive. In the following, we are able to answer completely this question: it depends on the sign of the coefficients. We recall that $A_3>0$. As argued above, if $A_2>0$, one must have $A_1<0$. However, if $A_2<0$, the sign of $A_1$ is not fixed. We prove the following result
\begin{lemma}
For $n=3$, the potential $V$ \eqref{potsingle} leads to a maximum (or bump) in $ -V'/V$ if and only if $A_2>0$.
\end{lemma}
\noindent We give in Appendix \ref{ap:proofs} the proof that there is always one, and only one bump for $A_2>0$, while there is none for $A_2<0$. This is illustrated by the graphs in Figure \ref{fig:n=3}.

For $n \geq 3$, we prove in Appendix \ref{ap:proofs} the following generalisation of the above
\begin{lemma2}
For $n \geq 3$, the potential $V$ \eqref{potsingle} leads to a maximum (or bump) in $ -V'/V$ if $A_{n-1}>0$.
\end{lemma2}
\noindent We recall $A_n>0$, and here, the sign of $A_{i<n-1}$ need not be specified (even though we know that one has to be negative).

This general analysis reveals the possibility of getting a bump in the ratio ${V'}^2/V^2$, and therefore in $\epsilon_V$, in case the former is a good approximation of the latter, for a purely exponential potential. This gives a first intuition, and possible explanation for what we observed for solution $m_{5577}^+ 4$ and its potential.

Using this analysis, a very first attempt to explain the bump would be through the dilaton alone: we noticed that the steepest descent had an important contribution from the dilaton. There are two reasons why we cannot conclude this way. First, as we will recall in the following, the dilaton alone admits an exponential potential of the type \eqref{LV}, with $n=3$. The signs of the coefficients, at the de Sitter critical point at least, are $A_1, A_3 >0, A_2<0$. According to Lemma 1, this would then not provide a bump so this is not an appropriate explanation. A second reason, however, is that the dilaton is stabilised at the critical point, so the above analysis does not apply since it assumes a maximum with $V''_0 <0$. One should then evaluate the sign of the coefficients at the maximum of $V(\phi)$, present between the minimum and the vanishing asymptotics. In the following, we rather consider the possibility of a two fields explanation.

\subsubsection{A two fields model}\label{sec:2fields}

As mentioned already, the most important contributions to the steepest descent trajectory for the  $m_{5577}^+ 4$ scalar potential are among the saxions. Since the dilaton alone may not explain the bump as it is stabilised at the critical point, the next simplest combination of saxions one can think of is the 4d dilaton $\tau$ and the 6d volume $\rho$: could those two be enough to reproduce the bump in $\epsilon_V$?

\begin{figure}[H]
\begin{center}
\includegraphics[width=0.75\textwidth]{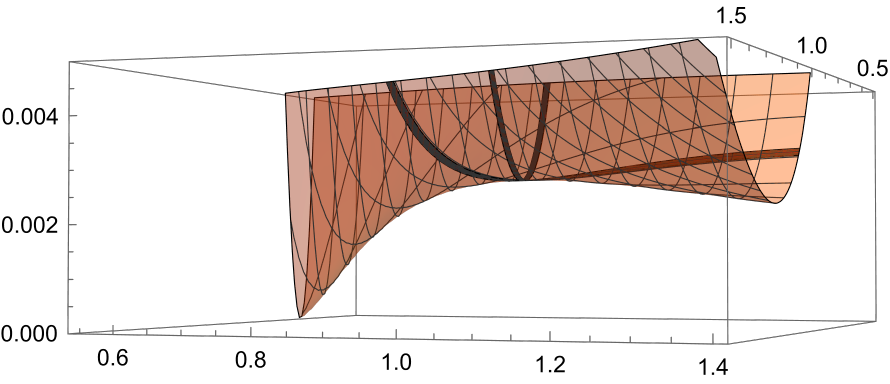}
\caption{$V(\rho,\tau)$ around solution $m_{5577}^+ 4$. We see that the potential de Sitter critical point is a maximum along a field direction, the tachyon. The dark, enlarged curves correspond to the potential along $\rho$ and along $\tau$: we see that the critical point is a minimum for both. As discussed in \cite[Sec. 3.3]{Andriot:2020wpp}, the tachyon then is due to the off-diagonal components of the Hessian.}\label{fig:3drhotau}
\end{center}
\end{figure}

It is well-known that the tachyon can already be found within the fields $(\rho,\tau, \sigma_I)$, where $\sigma_I$ are related to the internal volumes wrapped by the $O_5/D_5$ and $O_7/D_7$ sources. Sometimes, the tachyon can even be found within $(\rho,\tau)$ only, despite the fact that these two fields typically admit a minimum of the potential along their own profile. Interestingly, we verify that it is the case for $m_{5577}^+ 4$, and illustrate it in Figure \ref{fig:3drhotau}: a tachyon can be found considering only those two fields. This situation could then provide us with a simplified two fields model of the complete potential.

\begin{figure}[H]
\begin{center}
\includegraphics[width=0.75\textwidth]{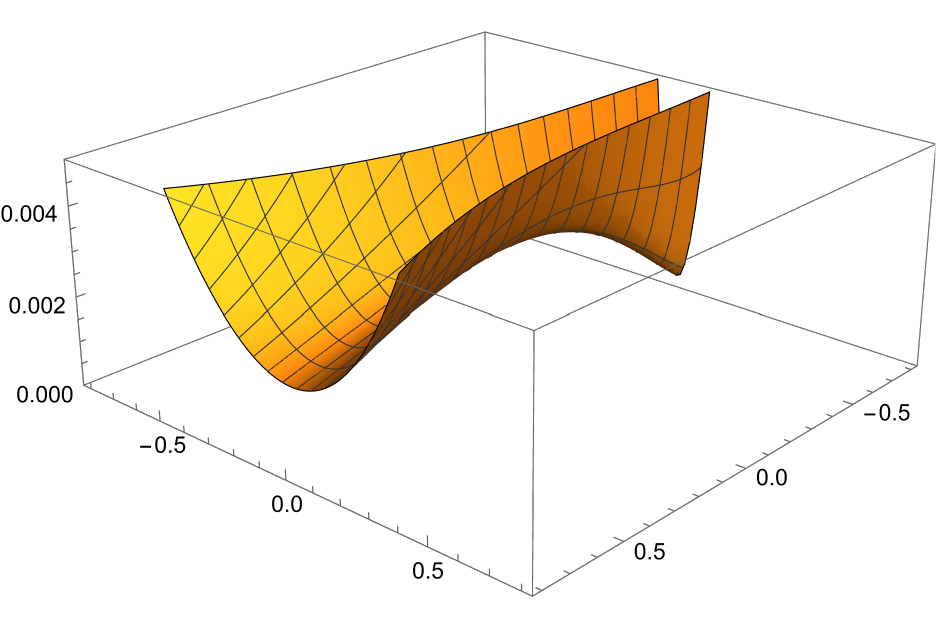}
\caption{$V(y,y_{\bot})$ around solution $m_{5577}^+ 4$. One verifies the orthonormality of the field basis $(y,y_{\bot})$, $y$ being the tachyonic direction.}\label{fig:3dyyb}
\end{center}
\end{figure}

We then start with the well-known \cite{Hertzberg:2007wc} two fields potential $V(\rho,\tau)$ \cite[(3.1)]{Andriot:2022yyj} around solution $m_{5577}^+ 4$, and $(\rho,\tau)$ field space metric. To study this two fields model, one can use and adapt for instance the code {\tt MSSSpec} \cite{Andriot:2022yyj} that provides the potential, the field space metric and the spectrum, giving useful numerical tools. We obtain the mass matrix in the canonical basis $(\hat{\rho},\hat{\tau})$ (see e.g.~\cite[(A.3)]{Andriot:2021rdy}), and its eigenvectors. The latter gives us the tachyonic field direction $y$ in a canonical basis, as well as the orthonormal direction $y_{\bot}$. One gets in particular the relation
\beq
\hat{\rho}= 0.35160\, y + 0.93615\, y_{\bot} \ ,\quad \hat{\tau} = 0.93615\, y -  0.35160\, y_{\bot} \ .
\eeq
The two fields potential then gets rewritten as
\bea
\hspace{-0.2in} \frac{V}{M_p^2}  = &  - \frac{1}{2}{\cal R}_6\  e^{-1.611 \, y - 0.26713 \, y_{\bot}} + \frac{1}{4} |H|^2 \ e^{-2.1852 \, y - 1.7959 \, y_{\bot}} - \frac{g_s T_{10}^{(5)}}{12} \ e^{-2.1294 \, y + 0.36367 \, y_{\bot}}  \label{potyyb} \\
& - \frac{g_s T_{10}^{(7)}}{16} e^{-1.8423 \, y + 1.1280 \, y_{\bot}} +\frac{1}{4} g_s^2 |F_1|^2 \ e^{-2.0737 \, y + 2.5232 \, y_{\bot}} + \frac{1}{4} g_s^2 |F_3|^2 \ e^{-2.6478 \, y + 0.99447 \, y_{\bot}} \ , \nn
\eea
with the solution data
\bea
& -\frac{1}{2} {\cal R}_6 = 0.016412 \ ,\ \frac{1}{4} |H|^2 = 0.017832 \ ,\ -\frac{g_s T_{10}^{(5)}}{12} = -0.062193 \ ,\\
& -\frac{g_s T_{10}^{(7)}}{16}= 0.0054428 \ ,\ \frac{1}{4} g_s^2 |F_1|^2 = 0.018045 \ ,\ \frac{1}{4} g_s^2 |F_3|^2 = 0.0073954 \ .\nn
\eea
This potential is illustrated in Figure \ref{fig:3dyyb}.

We first note that this potential \eqref{potyyb} is of the form \eqref{potsingle} for the tachyonic direction $y$: we have $n=6$, with $A_3<0$ and other coefficients positive. Therefore, according to Lemma 2, that direction should admit a bump, a point that we easily verify. If the steepest descent is close to the tachyonic direction, this could then provide an explanation.

We then run the steepest descent procedure on potential \eqref{potyyb}, starting at the critical point along the tachyonic direction. The corresponding notebook is {\tt Steepest descent 2 fields potential}. We consider 34 points, the last one giving $V \approx 5.8928 \cdot 10^{-8}$ and a field distance close to $8$. This allows a fair comparison to the results on the complete potential.

\begin{figure}[H]
\begin{center}
\begin{subfigure}[H]{0.45\textwidth}
\includegraphics[width=\textwidth]{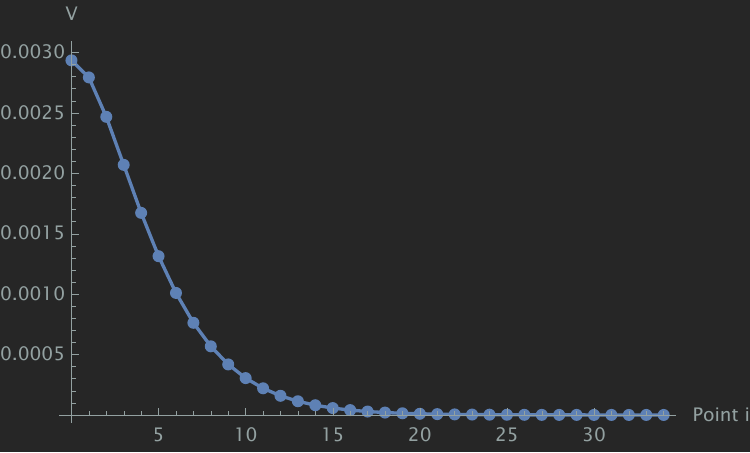}\caption{$V(y,y_{\bot})$}\label{fig:2fieldspot}
\end{subfigure}\qquad \quad
\begin{subfigure}[H]{0.45\textwidth}
\includegraphics[width=\textwidth]{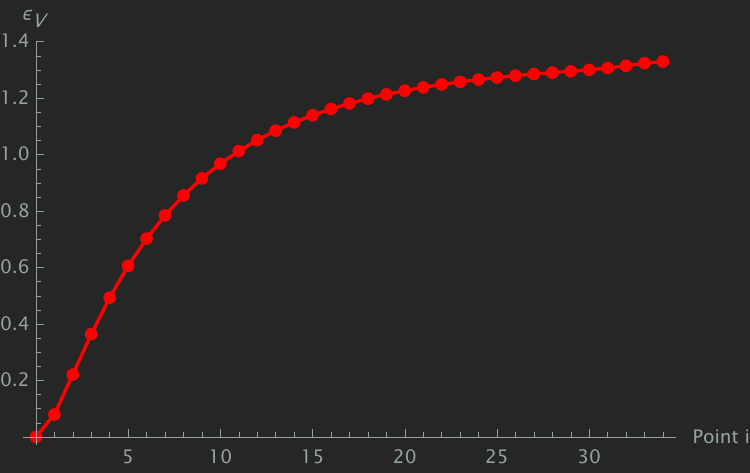}\caption{$\epsilon_V$}\label{fig:2fieldsepsilon}
\end{subfigure}\\
\begin{subfigure}[H]{0.45\textwidth}
\includegraphics[width=\textwidth]{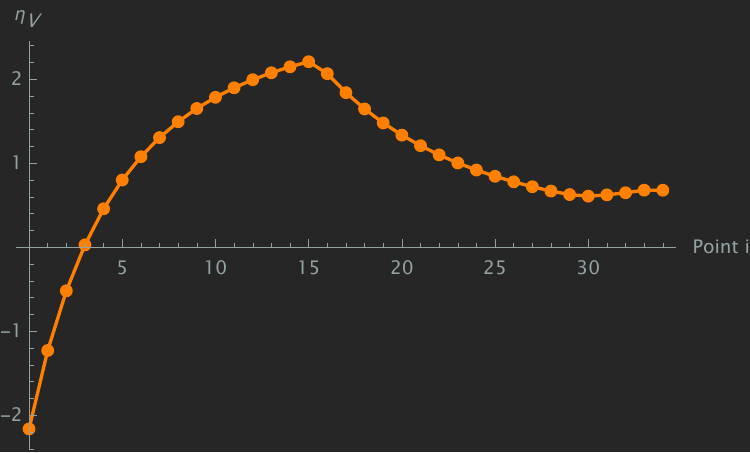}\caption{$\eta_V$}\label{fig:2fieldseta}
\end{subfigure}
\caption{Results of the steepest descent for the two fields potential $V(y,y_{\bot})$ \eqref{potyyb}.}\label{fig:2fields}
\end{center}
\end{figure}

The main result is that there is no bump in $\epsilon_V$, when following the steepest descent on this two fields model: see Figure \ref{fig:2fieldsepsilon}. Along this trajectory, the two fields start at 0 and end-up at $y=7.7795,\, y_{\bot}=0.38025$. This signals a small but clear deviation from the tachyon $y$ alone, hence preventing from benefiting from the tachyonic bump. This simplified model is then not enough to explain this feature of the complete potential.

\subsection{Linear contribution: the axion}\label{sec:axion}

We have seen that a pure exponential potential can lead to a bump in $\epsilon_V$ under certain circumstances. The simplified model with two fields, related to the dilaton and the volume, however did not admit such a bump. This is a first reason to consider going beyond purely exponentials. The complete potential also depends on axions, that rather have polynomial dependence; such a dependence could provide an origin to the bump. A second reason to think so is the comparison to the species scale. As discussed in Section \ref{sec:Ls}, the bump finds its origin there in a linear dependence in the field. It would then be interesting to see if the same type of dependence plays a role for the potential.

Along the steepest descent trajectory, all axions happen to take negligible values, except for the $B$-field \eqref{fields24}. Still, the $B$-field remains smaller than the saxions, but it may then be relevant to include it. The complete scalar potential is given in \cite[(2.21)]{Andriot:2022bnb}. For simplicity, we first set there the RR axions $C_k = 0$, as well as $g_{56}=0$, since these fields take much smaller values. Since the $B$-field axion takes small values (we consider its two components $b_{23}, b_{41}$), we could also keep its dependence up to the linear term. We keep it here up to the quadratic terms,\footnote{A quadratic term is not taken into account here for simplicity, proportional to $|F_3\w B_2|^2$. This is justified by the higher suppression due to the inverse metric factors.} to get the right behaviour when proceeding with the steepest descent, otherwise possibly falsifying the asymptotics getting for instance a negative potential. We then obtain the following simplified potential for solution $m_{5577}^+ 4$
\bea
\label{VIIB}
V & \approx \frac{M_p^2 }{2} \frac{e^{2\phi}}{vol_6} \bigg( -R_6 +\frac12 |H_3|^2 - e^{\phi}\sum_{p=5,7}\, \frac{T_{10}^{(p)}}{p+1} + \frac{e^{2\phi}}{2} \left( |F_1|^2 +|F_3|^2  \right) \\
& \qquad \qquad \qquad - b_{41} \left( H^{125}\,f^4{}_{25} + H^{345} \,f^1{}_{35} \right) - b_{23} \left( H^{125} \,f^3{}_{15} + H^{345} \,f^2{}_{45} \right)  \nn\\
& \qquad \qquad \qquad + e^{2\phi}\, b_{41} \left(-F_3^{145} F_{1\, 5} - F_3^{146} F_{1\, 6}\right) + e^{2\phi} b_{23} \left(F_3^{235} F_{1\, 5} + F_3^{236} F_{1\, 6}\right) \nn\\
& \qquad \qquad \qquad +\frac{1}{2} \left( b_{41} \,f^1{}_{35} + b_{23} \,f^2{}_{45} \right)^2  +\frac{1}{2} \left( b_{41} \,f^4{}_{25} + b_{23} \,f^3{}_{15} \right)^2 + \frac{e^{2\phi}}{2} (b_{41}^2 + b_{23}^2) |F_1|^2 \bigg)\nn
\eea
where the first line only has a dependence on the saxions ($\phi, g_{aa}$), referring to  \cite{Andriot:2022bnb} for details. The second and third lines are the approximated linear $B$-field axion potential.\footnote{Note that linear terms in the $B$-field axion reproduce the $B$-field equation of motion (at the critical point), given here in the smeared limit by $d *_6\! H = g_s^2 F_1 \w *_6 F_3$. Away from the critical point, one may neglect the RR terms because of the dilaton factors.} The squares in the last line involve 3 inverse metric components, as in the lines before where we lifted flux indices. Fluxes in that expression are the background ones.

We now consider that the trajectory can also be along the $B$-field axion, and the latter is (almost) canonically normalized, so we now get a linear term in that field direction, on top of the usual exponentials; as argued above, we may neglect the quadratic terms in the analysis. This form of the potential is very reminiscent of the species scale expression \eqref{Lsapprox}. We can make the comparison more precise: one can verify that $H^{125}\,f^4{}_{25} + H^{345} \,f^1{}_{35}>0$ and $H^{125} \,f^3{}_{15} + H^{345} \,f^2{}_{45} >0$, and that along the trajectory, $b_{41}>0$ and $b_{23}>0$. The third line becomes smaller and smaller due to the dilaton factor; we still note that its signs are opposite with $-F_3^{145} F_{1\, 5} - F_3^{146} F_{1\, 6}>0$ and $F_3^{235} F_{1\, 5} + F_3^{236} F_{1\, 6}>0$. So the coefficients of the linear terms are overall {\sl negative}, thanks to the second line. This situation is the contrary to the species scale expression \eqref{Lsapprox}. Note also that the first line in \eqref{VIIB} gives the whole potential at the critical point, so it is there, and probably beyond, positive, as for the first term in \eqref{Lsapprox}. Due to the difference in the linear term sign, a bump may not appear here.\\

To verify this, we add this $B$-field contribution to the two fields model considered in Section \ref{sec:2fields}; the latter is captured by the first line of the potential \eqref{VIIB}. To get a model including the $B$-field, we need the axion kinetic term. Following \cite{Andriot:2022bnb}, it is given by
\beq
- \frac{1}{2} \int \d^4 x \sqrt{|g_4|}\ \frac{M_p^2}{2} |\del B_2|^2 = - \frac{1}{2} \int \d^4 x \sqrt{|g_4|}\ \frac{M_p^2}{2} \rho^{-2} \left( (\del b_{14})^2 +  (\del b_{23})^2 \right) \ ,
\eeq
where we use the fluctuation $\rho$ along an internal diagonal metric, with the background values being $g_{11}=g_{22}=g_{33}=g_{44}=1$. We deduce the field space metric for the 2 $B$-field axions $g_{BB}=\frac{M_p^2}{2 \rho^{2}} $, within the set of fields considered. Finally, the internal metric entering the $B$-field linear terms in the potential give rise to a $\rho^{-3}$ factor. Expressing $\rho$ in terms of $y, y_{\bot}$, we then have a model depending on the four fields of interest, $(y,y_{\bot},b_{23}, b_{41})$, on which we can proceed with the steepest descent.\footnote{Note that the quadratic terms slightly modify the mass spectrum, in particular the tachyonic direction from which we are starting. We neglect here this modification, and still launch the steepest descent along the same tachyonic direction as before; the minimization procedure could a priori rectify the trajectory if necessary.}

We do so in the notebook {\tt Steepest descent 2 fields potential with axions}. We consider 34 points, the last one giving $V \approx 6.4269 \cdot 10^{-8}$. The values reached by the fields, $y = 7.7245,\, y_{\bot} = 0.33862,\, b_{41} = 0.16998,\, b_{23} = 0.48576$, are close to those of the two fields model, so we should again get a fair comparison to the results of the complete potential \eqref{fields24}. We note also that the $B$-field values are of the same order as those reached in the complete potential. The fact those are small, as assumed, but non-trivial, underlines the relevance of these axions.

\begin{figure}[H]
\begin{center}
\begin{subfigure}[H]{0.45\textwidth}
\includegraphics[width=\textwidth]{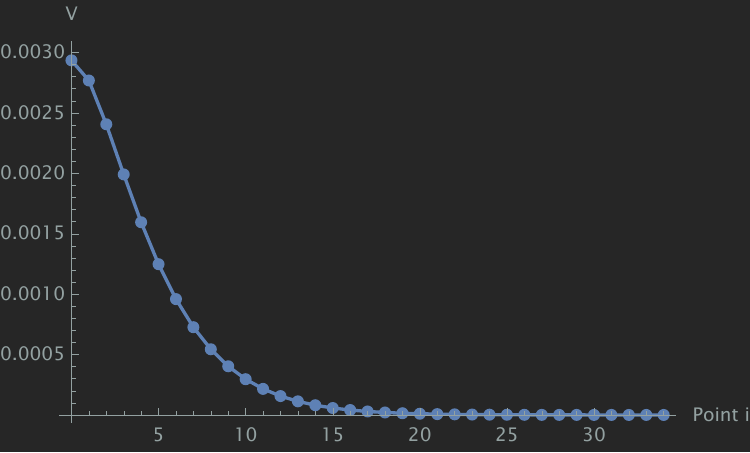}\caption{$V(y,y_{\bot},b_{23}, b_{41})$}\label{fig:2fieldsaxionpot}
\end{subfigure}\qquad \quad
\begin{subfigure}[H]{0.45\textwidth}
\includegraphics[width=\textwidth]{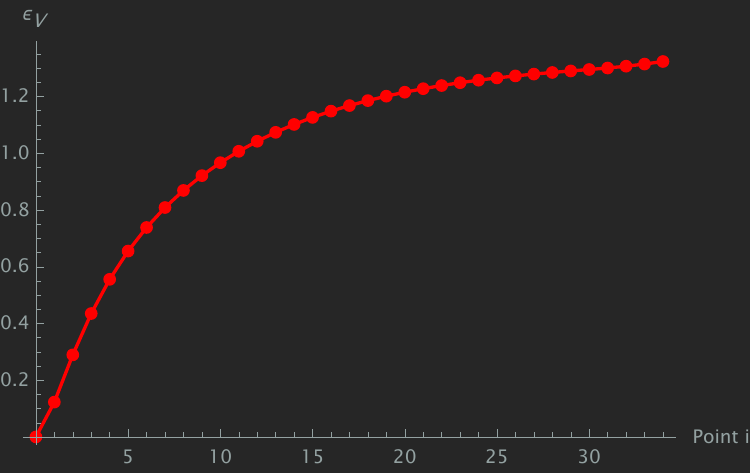}\caption{$\epsilon_V$}\label{fig:2fieldsaxionepsilon}
\end{subfigure}
\caption{Results of the steepest descent for the four fields potential $V(y,y_{\bot},b_{23}, b_{41})$ \eqref{VIIB}.}\label{fig:2fieldsaxion}
\end{center}
\end{figure}

The result of the steepest descent in this simplified model is an absence of bump in $\epsilon_V$ (see Figure \ref{fig:2fieldsaxionepsilon}), as anticipated from the analysis of the linear term sign; we note that there is actually very little change with or without the axion. So the intuition of species scale and its linear term fails here.

\subsection{More exponentials}\label{sec:moreexp}

The previous analysis has shown that axions, giving linear dependence in the potential, are not responsible for the observed bump in $\epsilon_V$ along the steepest descent. This is to be contrasted with the species scale where the linear term gives an origin to that maximum. It was also noticed that the linear term in the potential has the opposite sign w.r.t. to the species scale, hence the difference.

If the axions play no role for the potential, then we are back to pure exponential dependence. We have seen in Section \ref{sec:expan} with Lemma 1 and 2 that having only exponentials can generate a bump. The simplified model with two fields, the volume and the dilaton, appeared however not enough to reproduce the bump. We then decide to consider a model where we include all saxions, ${\phi,g_{aa}}$, i.e.~remove the 7 axions. More precisely, from the complete potential and field space metric, we set the 7 axions to their critical point value, which is 0, and consider the restricted 7 fields resulting model. We recompute the tachyonic direction in that case. We then proceed with the steepest descent.

We do so in the notebook {\tt Steepest descent without 7 axions}. We start at the extremum with $g_{aa}=1, \, \phi =0$. We consider 30 points, ending up with $V \approx 8.4910 \cdot 10^{-9}$ at the point $g_{11} = 2.29273,\, g_{22} = 1.49132,\, g_{33} = 2.0258,\, g_{44} = 2.44493,\, g_{55}= 3.19269,\, g_{66} = 2.06666,\, \phi= -5.58251$. This is close enough to the complete potential \eqref{fields24} to offer a fair comparison.

\begin{figure}[H]
\begin{center}
\begin{subfigure}[H]{0.45\textwidth}
\includegraphics[width=\textwidth]{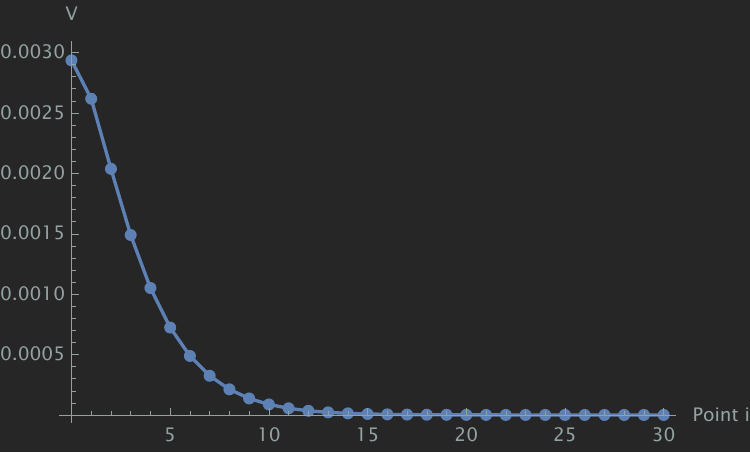}\caption{$V(\phi,g_{aa})$}\label{fig:without7axionspot}
\end{subfigure}\qquad \quad
\begin{subfigure}[H]{0.45\textwidth}
\includegraphics[width=\textwidth]{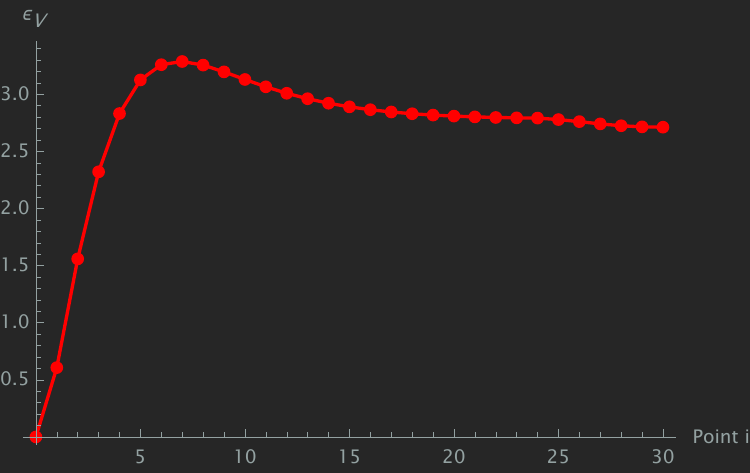}\caption{$\epsilon_V$}\label{fig:without7axionsepsilon}
\end{subfigure}
\caption{Results of the steepest descent for the seven fields (saxions) potential.}\label{fig:without7axions}
\end{center}
\end{figure}

The result of the steepest descent in this simplified model with only saxions is a bump in the $\epsilon_V$ curve: see Figure \ref{fig:without7axionsepsilon}.

This confirms that axions are not relevant here, and it provides to us an origin to the bump observed with the complete potential. {\sl The bump comes from a purely exponential potential generated by the saxions}. Interestingly, the two fields model was only a simplified version of the former, but was not enough. The volume, which is the universal diagonal fluctuation, is then not sufficient and one needs the further variations of each radius $g_{aa}$. It would be interesting to work out in more details the potential expression in terms of exponentials, and trace the origin to a property like Lemma 2. The 7 fields, with an involved field space metric (a full square), make any such attempt much more complicated than in the two fields model. Still, the fact that it is purely exponential, and results like Lemma 2, give us a good understanding of the origin of this bump; enough to see that it is actually very different than that of the species scale.

\section{Outlook}\label{sec:outlook}

In this work, considering a quantum gravity EFT \eqref{EFTV}, we are interested in the comparison of the scalar potential and the species scale, and their field-dependent features. This is motivated by the two inequalities \eqref{LV} and \eqref{ratecompare}, that bound the potential by the species scale, and order their rates accordingly. A first task has been to find a concrete scalar potential from a string compactification that has the following features: being positive, admitting a de Sitter critical point in the bulk of field space, with (at least) one tachyonic direction, and having this field direction asymptoting to zero. Indeed, these features are those of the species scale as displayed in Figure \ref{fig:Ls}. We explain in the Introduction and in Section \ref{sec:m55774} that this is already a non-trivial task. We were led to consider the de Sitter solution $m_{5577}^+ 4$, its 14-fields scalar potential and the field trajectory following the steepest descent, giving us one example with the desired behaviour. Doing so, we then observed a bump in the potential rate (or $\epsilon_V$), i.e.~a local maximum, another feature of the species scale shown in Figure \ref{fig:Lsratio}. We investigate in Section \ref{sec:bump} the origin of that bump for the scalar potential. We conclude that it is due to the dependence on the dilaton and the radii (or diagonal metric fluctuations $g_{aa}$), but that the dilaton and overall volume alone are not enough to explain it. Those fields, when canonically normalized, enter as exponentials in the scalar potential; the bump then has a different origin as the one of the species scale, which was due to a linear field dependence. We verify in particular that the axions, that could provide a linear dependence in the scalar potential, do not generate the bump here. These differences are interesting and indicate that the inequalities \eqref{LV} and \eqref{ratecompare} are in the end not as tight as they could be; if the origin of the bump was similar, one could have imagined a saturation of these inequalities.\\

An important loophole in the above is however that the compactifications considered are different. It is not clear whether the features observed for the species scale, obtained on a Calabi-Yau compactification, versus those of the scalar potential, obtained a group manifold flux compactification, are actually comparable and go beyond their respective compactification framework. A first difference is that the former admits no scalar potential. The latter however should admit a species scale, but we have not determined it. It would be interesting to do so, following the idea that it appears in higher derivative corrections. Since it is a flux compactification, one should however understand which higher derivative correction(s) to consider (gravitational and/or fluxes?). At first sight, one would obtain a combination of exponentials of the dilaton and radii, which might be comparable to the scalar potential.

A second difference is that most species scale examples in \cite{vandeHeisteeg:2023ubh} are single field. A fair comparison to a scalar potential would be a multifield situation. In that case, what field direction and asymptotics should be chosen to study the evolution of the species scale? Again, picking the steepest descent trajectory would make sense as it would bring the tightest constraints and quantum gravity bounds, thanks to a minimal species scale. It would then be interesting to see whether the features observed so far for the species scale would survive, or become direction dependent, as studied here for the scalar potential.

\vfill

\subsection*{Acknowledgements}

We thank Tom Rudelius for collaboration at an early stage of this project. We thank D.~van de Heisteeg, C.~Vafa and M.~Wiesner for insightful exchanges on this topic. We thank Harvard University and the CMSA, as well as Utrecht University, and their members, for warm hospitality during the completion of this project.

\newpage

\begin{appendix}

\section{Proofs for an exponential potential}\label{ap:proofs}

In this appendix, we prove Lemma 1 and 2 discussed in Section \ref{sec:expan}, that we repeat here and in the following for convenience
\begin{lemma}
For $n=3$, the potential $V$ \eqref{potsingle} leads to a maximum (or bump) in $ -V'/V$ if and only if $A_2>0$.
\end{lemma}
To prove Lemma 1, we use expressions \eqref{Aiexp} of $A_{1,2,3}$, we recall that $V_0>0$ and ${V''}_0<0$. We finally recall that $A_3>0$, and that if $A_2>0$ then $A_1<0$.

\begin{proof}
We start with the case $A_2 > 0$, for which we show that there is always a bump. Since $A_3>0$, $A_2>0$ and $A_1<0$, one has from \eqref{bla} the following expression
\beq
{V'}^2 - V\, V'' = X_2 X_3 \left( (a_1 - a_2)^2 \left|\frac{A_1}{A_3}\right| e^{-(a_3-a_1) \hat{\varphi}} + (a_1 - a_3)^2 \left|\frac{A_1}{A_2}\right| e^{-(a_2-a_1) \hat{\varphi}} - (a_2 - a_3)^2 \right) \ .
\eeq
Let us rewrite the right-hand side as $X_2 X_3 (f(\hat{\varphi}) - C^2)$ to simplify the discussion.
The first two terms giving $f(\hat{\varphi})$ are positive and the last one, $- C^2$, is negative. At $\hat{\varphi}=0$, we know that ${V'}^2 - V\, V''$ is positive, so $f(0) - C^2 >0$. When $\hat{\varphi} \rightarrow \infty$, $f(\hat{\varphi}) \rightarrow 0$, so $f(\infty) - C^2 <0$. Since $f(\hat{\varphi})$ is a continuous and monotonic function, we conclude there always exists one, and only one, value $\hat{\varphi}_b > 0$ such that $f(\hat{\varphi}_b) - C^2 = 0$, i.e.~${V'}^2 - V\, V''\, |_{\hat{\varphi}_b} = 0$. We conclude that the ratio of interest $|V'|/V $ admits a single bump whenever $A_2 > 0$.

We turn to $A_2<0$, for which $A_1$ can have either sign. One gets
\beq
{V'}^2 - V\, V'' = |X_2 X_3| \left(A_1 \left( (a_1 - a_2)^2 \frac{1}{|A_3|} e^{-(a_3-a_1) \hat{\varphi}} - (a_1 - a_3)^2 \frac{1}{|A_2|} e^{-(a_2-a_1) \hat{\varphi}} \right) + (a_2 - a_3)^2 \right) \ .
\eeq
Using expressions \eqref{Aiexp}, denoting by $N_i$ the numerators of these expressions, and using the signs $A_3>0, A_2<0$, one can show
\beq
A_1 \left( (a_1 - a_2)^2 \frac{1}{|A_3|}  - (a_1 - a_3)^2 \frac{1}{|A_2|} \right) = -(a_3-a_2)^2  \frac{N_1}{|N_2 N_3|}  (a_1^2 V_0 + {V''}_0 ) \ .
\eeq
One has $a_1^2 V_0 + {V''}_0  > N_3 >0$. For $A_1<0$, one concludes that the above is positive, while it is negative for $A_1>0$. Therefore, one has
\beq
A_1<0:\ - (a_1 - a_3)^2 \frac{A_1}{|A_2|} > - (a_1 - a_2)^2 \frac{A_1}{|A_3|} > - (a_1 - a_2)^2 \frac{A_1}{|A_3|}\, \frac{e^{-(a_3-a_1) \hat{\varphi}}}{e^{-(a_2-a_1) \hat{\varphi}}} \ ,
\eeq
from which we conclude that ${V'}^2 - V\, V'' > 0$ for $A_1<0$, and there is then no bump. Furthermore, we get
\beq
A_1>0:\ - (a_1 - a_3)^2 \frac{A_1}{|A_2|} < - (a_1 - a_2)^2 \frac{A_1}{|A_3|} < - (a_1 - a_2)^2 \frac{A_1}{|A_3|}\, \frac{e^{-(a_3-a_1) \hat{\varphi}}}{e^{-(a_2-a_1) \hat{\varphi}}} \ ,
\eeq
so the following function is negative
\beq
g(\hat{\varphi}) = A_1 \left( (a_1 - a_2)^2 \frac{1}{|A_3|} e^{-(a_3-a_1) \hat{\varphi}} - (a_1 - a_3)^2 \frac{1}{|A_2|} e^{-(a_2-a_1) \hat{\varphi}} \right) <0 \ .
\eeq
The first term has a lower amplitude but a higher (faster) exponential rate than the second one. One can then see graphically that the difference between the two, i.e.~the function $g$, admits a negative minimum. That minimum is reached at a value $\hat{\varphi}_0$ such that $g'(\hat{\varphi}_0)=0$, corresponding to
\beq
(a_1 - a_2)^2 \frac{1}{|A_3|} e^{-(a_3-a_1) \hat{\varphi}_0} = \frac{a_2-a_1}{a_3-a_1} (a_1 - a_3)^2 \frac{1}{|A_2|} e^{-(a_2-a_1) \hat{\varphi}_0} \ .
\eeq
We deduce in that case, with $A_2<0$ and $A_1>0$, that
\bea
\frac{{V'}^2 - V\, V''}{|X_2 X_3|}  & \geq (a_2 - a_3)^2 + g(\hat{\varphi}_0) \\
&  \geq (a_2 - a_3)^2 \left( 1 - \frac{N_1}{|N_2|}  \right) \nn\\
&  = \frac{(a_2 - a_3)^2}{|N_2|} a_3 (a_1 - a_2) V_0 > 0 \nn \ ,
\eea
where we used \eqref{Aiexp}. We conclude again that ${V'}^2 - V\, V'' > 0$, now for $A_1>0$, and there is no bump.
\end{proof}

We turn to
\begin{lemma2}
For $n \geq 3$, the potential $V$ \eqref{potsingle} leads to a maximum (or bump) in $ -V'/V$ if $A_{n-1}>0$.
\end{lemma2}

\begin{proof}
Using \eqref{bla}, we can write
\bea
& {V'}^2 - V V'' = X_{n-1} X_{n} \left( f(\hat{\varphi}) - (a_{n-1} - a_n)^2 \right) \ ,\\
{\rm where}\  & f(\hat{\varphi}) = - \sum_{i< j< n-1} (a_i -a_j)^2 \frac{A_i A_j}{A_{n-1} A_n} e^{-(a_n-a_i) \hat{\varphi}}\, e^{-(a_{n-1}-a_j) \hat{\varphi}} \\
&\phantom{f(\hat{\varphi}) =}  - \!\! \sum_{{\scriptsize \begin{array}{c} i < n-1,\\ j= n-1,n,\\ j_{\bot}=n, n-1 \end{array}}} \!\! (a_i - a_j)^2 \frac{A_i}{A_{j_{\bot}}}\, e^{-(a_{j_{\bot}} - a_i) \hat{\varphi}} \ .
\eea
We consider $A_{n-1}>0$, therefore $X_{n-1} X_n >0$. By definition, we know that ${V'}^2 - V V''>0$ at $\hat{\varphi}=0$, and so is $({V'}^2 - V V'')/(X_{n-1} X_n) $. In addition, we verify that $f \rightarrow 0$ for $\hat{\varphi} \rightarrow \infty$, so $({V'}^2 - V V'')/(X_{n-1} X_n) < 0 $ at $\hat{\varphi} \rightarrow \infty$. Since $f$ is a continuous function, we deduce that $({V'}^2 - V V'')/(X_{n-1} X_n) $ must vanish at least once for a positive $\hat{\varphi}$, and so does ${V'}^2 - V V''$. Given the signs, this corresponds to a maximum of $-V'/V$, so this ratio then admits a bump.
\end{proof}

\end{appendix}

\newpage

\providecommand{\href}[2]{#2}\begingroup\raggedright\endgroup


\begin{thebibliography}{10}

\bibitem{Vafa:2005ui}
C.~Vafa, \emph{{The String landscape and the swampland}},
  [\href{https://arxiv.org/abs/hep-th/0509212}{{\ttfamily hep-th/0509212}}].

\bibitem{Palti:2019pca}
E.~Palti, \emph{{The Swampland: Introduction and Review}},
  \href{https://doi.org/10.1002/prop.201900037}{\emph{Fortsch. Phys.}
  {\bfseries 67} (2019) 1900037}
  [\href{https://arxiv.org/abs/1903.06239}{{\ttfamily 1903.06239}}].

\bibitem{vanBeest:2021lhn}
M.~van Beest, J.~Calder\'on-Infante, D.~Mirfendereski and I.~Valenzuela,
  \emph{{Lectures on the Swampland Program in String Compactifications}},
  \href{https://doi.org/10.1016/j.physrep.2022.09.002}{\emph{Phys. Rept.}
  {\bfseries 989} (2022) 1} [\href{https://arxiv.org/abs/2102.01111}{{\ttfamily
  2102.01111}}].

\bibitem{Agmon:2022thq}
N.B.~Agmon, A.~Bedroya, M.J.~Kang and C.~Vafa, \emph{{Lectures on the string
  landscape and the Swampland}},
  [\href{https://arxiv.org/abs/2212.06187}{{\ttfamily 2212.06187}}].

\bibitem{Dvali:2007hz}
G.~Dvali, \emph{{Black Holes and Large N Species Solution to the Hierarchy
  Problem}}, \href{https://doi.org/10.1002/prop.201000009}{\emph{Fortsch.
  Phys.} {\bfseries 58} (2010) 528}
  [\href{https://arxiv.org/abs/0706.2050}{{\ttfamily 0706.2050}}].

\bibitem{Dvali:2007wp}
G.~Dvali and M.~Redi, \emph{{Black Hole Bound on the Number of Species and
  Quantum Gravity at LHC}},
  \href{https://doi.org/10.1103/PhysRevD.77.045027}{\emph{Phys. Rev. D}
  {\bfseries 77} (2008) 045027}
  [\href{https://arxiv.org/abs/0710.4344}{{\ttfamily 0710.4344}}].

\bibitem{Dvali:2009ks}
G.~Dvali and D.~Lust, \emph{{Evaporation of Microscopic Black Holes in String
  Theory and the Bound on Species}},
  \href{https://doi.org/10.1002/prop.201000008}{\emph{Fortsch. Phys.}
  {\bfseries 58} (2010) 505} [\href{https://arxiv.org/abs/0912.3167}{{\ttfamily
  0912.3167}}].

\bibitem{Dvali:2010vm}
G.~Dvali and C.~Gomez, \emph{{Species and Strings}},
  [\href{https://arxiv.org/abs/1004.3744}{{\ttfamily 1004.3744}}].

\bibitem{Dvali:2012uq}
G.~Dvali, C.~Gomez and D.~Lust, \emph{{Black Hole Quantum Mechanics in the
  Presence of Species}},
  \href{https://doi.org/10.1002/prop.201300002}{\emph{Fortsch. Phys.}
  {\bfseries 61} (2013) 768} [\href{https://arxiv.org/abs/1206.2365}{{\ttfamily
  1206.2365}}].

\bibitem{Castellano:2021mmx}
A.~Castellano, A.~Herr\'aez and L.E.~Ib\'a\~nez, \emph{{IR/UV mixing, towers of
  species and swampland conjectures}},
  \href{https://doi.org/10.1007/JHEP08(2022)217}{\emph{JHEP} {\bfseries 08}
  (2022) 217} [\href{https://arxiv.org/abs/2112.10796}{{\ttfamily
  2112.10796}}].

\bibitem{Long:2021jlv}
C.~Long, M.~Montero, C.~Vafa and I.~Valenzuela, \emph{{The desert and the
  swampland}}, \href{https://doi.org/10.1007/JHEP03(2023)109}{\emph{JHEP}
  {\bfseries 03} (2023) 109}
  [\href{https://arxiv.org/abs/2112.11467}{{\ttfamily 2112.11467}}].

\bibitem{Castellano:2022bvr}
A.~Castellano, A.~Herr\'aez and L.E.~Ib\'a\~nez, \emph{{The Emergence Proposal
  in Quantum Gravity and the Species Scale}},
  [\href{https://arxiv.org/abs/2212.03908}{{\ttfamily 2212.03908}}].

\bibitem{vandeHeisteeg:2022btw}
D.~van~de Heisteeg, C.~Vafa, M.~Wiesner and D.H.~Wu, \emph{{Moduli-dependent
  Species Scale}},  [\href{https://arxiv.org/abs/2212.06841}{{\ttfamily
  2212.06841}}].

\bibitem{Cribiori:2022nke}
N.~Cribiori, D.~Lust and G.~Staudt, \emph{{Black hole entropy and
  moduli-dependent species scale}},
  [\href{https://arxiv.org/abs/2212.10286}{{\ttfamily 2212.10286}}].

\bibitem{vandeHeisteeg:2023ubh}
D.~van~de Heisteeg, C.~Vafa and M.~Wiesner, \emph{{Bounds on Species Scale and
  the Distance Conjecture}},
  [\href{https://arxiv.org/abs/2303.13580}{{\ttfamily 2303.13580}}].

\bibitem{Ooguri:2006in}
H.~Ooguri and C.~Vafa, \emph{{On the Geometry of the String Landscape and the
  Swampland}},
  \href{https://doi.org/10.1016/j.nuclphysb.2006.10.033}{\emph{Nucl. Phys. B}
  {\bfseries 766} (2007) 21}
  [\href{https://arxiv.org/abs/hep-th/0605264}{{\ttfamily hep-th/0605264}}].

\bibitem{Klaewer:2016kiy}
D.~Klaewer and E.~Palti, \emph{{Super-Planckian Spatial Field Variations and
  Quantum Gravity}}, \href{https://doi.org/10.1007/JHEP01(2017)088}{\emph{JHEP}
  {\bfseries 01} (2017) 088}
  [\href{https://arxiv.org/abs/1610.00010}{{\ttfamily 1610.00010}}].

\bibitem{Baume:2016psm}
F.~Baume and E.~Palti, \emph{{Backreacted Axion Field Ranges in String
  Theory}}, \href{https://doi.org/10.1007/JHEP08(2016)043}{\emph{JHEP}
  {\bfseries 08} (2016) 043}
  [\href{https://arxiv.org/abs/1602.06517}{{\ttfamily 1602.06517}}].

\bibitem{Lee:2019wij}
S.-J.~Lee, W.~Lerche and T.~Weigand, \emph{{Emergent strings from infinite
  distance limits}}, \href{https://doi.org/10.1007/JHEP02(2022)190}{\emph{JHEP}
  {\bfseries 02} (2022) 190}
  [\href{https://arxiv.org/abs/1910.01135}{{\ttfamily 1910.01135}}].

\bibitem{Obied:2018sgi}
G.~Obied, H.~Ooguri, L.~Spodyneiko and C.~Vafa, \emph{{De Sitter Space and the
  Swampland}},  [\href{https://arxiv.org/abs/1806.08362}{{\ttfamily
  1806.08362}}].

\bibitem{Rudelius:2019cfh}
T.~Rudelius, \emph{{Conditions for (No) Eternal Inflation}},
  \href{https://doi.org/10.1088/1475-7516/2019/08/009}{\emph{JCAP} {\bfseries
  08} (2019) 009} [\href{https://arxiv.org/abs/1905.05198}{{\ttfamily
  1905.05198}}].

\bibitem{Bedroya:2019snp}
A.~Bedroya and C.~Vafa, \emph{{Trans-Planckian Censorship and the Swampland}},
  \href{https://doi.org/10.1007/JHEP09(2020)123}{\emph{JHEP} {\bfseries 09}
  (2020) 123} [\href{https://arxiv.org/abs/1909.11063}{{\ttfamily
  1909.11063}}].

\bibitem{Andriot:2022brg}
D.~Andriot, L.~Horer and G.~Tringas, \emph{{Negative scalar potentials and the
  swampland: an Anti-Trans-Planckian Censorship Conjecture}},
  \href{https://doi.org/10.1007/JHEP04(2023)139}{\emph{JHEP} {\bfseries 04}
  (2023) 139} [\href{https://arxiv.org/abs/2212.04517}{{\ttfamily
  2212.04517}}].

\bibitem{Hebecker:2018vxz}
A.~Hebecker and T.~Wrase, \emph{{The Asymptotic dS Swampland Conjecture - a Simplified Derivation and a Potential Loophole}},   \href{https://doi.org/10.1002/prop.201800097}{\emph{Fortsch. Phys.} \textbf{67} (2019) no.1-2 1800097} [\href{https://arxiv.org/abs/1810.08182}{{\ttfamily 1810.08182}}].

\bibitem{Scalisi:2018eaz}
M.~Scalisi and I.~Valenzuela, \emph{{Swampland distance conjecture, inflation
  and $\alpha$-attractors}},
  \href{https://doi.org/10.1007/JHEP08(2019)160}{\emph{JHEP} {\bfseries 08}
  (2019) 160} [\href{https://arxiv.org/abs/1812.07558}{{\ttfamily
  1812.07558}}].

\bibitem{Andriot:2022way}
D.~Andriot, L.~Horer and P.~Marconnet, \emph{{Charting the landscape of (anti-)
  de Sitter and Minkowski solutions of 10d supergravities}},
  \href{https://doi.org/10.1007/JHEP06(2022)131}{\emph{JHEP} {\bfseries 06}
  (2022) 131} [\href{https://arxiv.org/abs/2201.04152}{{\ttfamily
  2201.04152}}].

\bibitem{Bena:2023sks}
I.~Bena, M.~Gra\~na and T.~Van~Riet, \emph{{Trustworthy de Sitter
  compactifications of string theory: a comprehensive review}},
  \href{https://arxiv.org/abs/2303.17680}{{\ttfamily 2303.17680}}.

\bibitem{Giddings:2001yu}
S.B.~Giddings, S.~Kachru and J.~Polchinski, \emph{{Hierarchies from fluxes in
  string compactifications}},
  \href{https://doi.org/10.1103/PhysRevD.66.106006}{\emph{Phys. Rev. D}
  {\bfseries 66} (2002) 106006}
  [\href{https://arxiv.org/abs/hep-th/0105097}{{\ttfamily hep-th/0105097}}].

\bibitem{Calderon-Infante:2022nxb}
J.~Calder\'on-Infante, I.~Ruiz and I.~Valenzuela, \emph{{Asymptotic Accelerated
  Expansion in String Theory and the Swampland}},
  [\href{https://arxiv.org/abs/2209.11821}{{\ttfamily 2209.11821}}].

\bibitem{Andriot:2017jhf}
D.~Andriot, \emph{{On classical de Sitter and Minkowski solutions with
  intersecting branes}},
  \href{https://doi.org/10.1007/JHEP03(2018)054}{\emph{JHEP} {\bfseries 03}
  (2018) 054} [\href{https://arxiv.org/abs/1710.08886}{{\ttfamily
  1710.08886}}].

\bibitem{Andriot:2022yyj}
D.~Andriot, L.~Horer and P.~Marconnet, \emph{{Exploring the landscape of
  (anti-) de Sitter and Minkowski solutions: group manifolds, stability and
  scale separation}},
  \href{https://doi.org/10.1007/JHEP08(2022)109}{\emph{JHEP} {\bfseries 08}
  (2022) 109} [\href{https://arxiv.org/abs/2204.05327}{{\ttfamily
  2204.05327}}].

\bibitem{Andriot:2022bnb}
D.~Andriot, P.~Marconnet, M.~Rajaguru and T.~Wrase, \emph{{Automated consistent
  truncations and stability of flux compactifications}}, \href{https://doi.org/10.1007/JHEP12(2022)026}{\emph{JHEP} {\bfseries 12}
  (2022) 026} [\href{https://arxiv.org/abs/2209.08015}{{\ttfamily 2209.08015}}].

\bibitem{Banlaki:2018ayh}
A.~Banlaki, A.~Chowdhury, C.~Roupec and T.~Wrase, \emph{{Scaling limits of dS
  vacua and the swampland}},
  \href{https://doi.org/10.1007/JHEP03(2019)065}{\emph{JHEP} {\bfseries 03}
  (2019) 065} [\href{https://arxiv.org/abs/1811.07880}{{\ttfamily
  1811.07880}}].

\bibitem{Junghans:2018gdb}
D.~Junghans, \emph{{Weakly Coupled de Sitter Vacua with Fluxes and the
  Swampland}}, \href{https://doi.org/10.1007/JHEP03(2019)150}{\emph{JHEP}
  {\bfseries 03} (2019) 150}
  [\href{https://arxiv.org/abs/1811.06990}{{\ttfamily 1811.06990}}].

\bibitem{Andriot:2019wrs}
D.~Andriot, \emph{{Open problems on classical de Sitter solutions}},
  \href{https://doi.org/10.1002/prop.201900026}{\emph{Fortsch. Phys.}
  {\bfseries 67} (2019) 1900026}
  [\href{https://arxiv.org/abs/1902.10093}{{\ttfamily 1902.10093}}].

\bibitem{Grimm:2019ixq}
T.W.~Grimm, C.~Li and I.~Valenzuela, \emph{{Asymptotic Flux Compactifications
  and the Swampland}},
  \href{https://doi.org/10.1007/JHEP06(2020)009}{\emph{JHEP} {\bfseries 06}
  (2020) 009} [\href{https://arxiv.org/abs/1910.09549}{{\ttfamily
  1910.09549}}].

\bibitem{Andriot:2020vlg}
D.~Andriot, P.~Marconnet and T.~Wrase, \emph{{Intricacies of classical de
  Sitter string backgrounds}},
  \href{https://doi.org/10.1016/j.physletb.2020.136015}{\emph{Phys. Lett. B}
  {\bfseries 812} (2021) 136015}
  [\href{https://arxiv.org/abs/2006.01848}{{\ttfamily 2006.01848}}].

\bibitem{Dine:1985he}
M.~Dine and N.~Seiberg, \emph{{Is the Superstring Weakly Coupled?}},
  \href{https://doi.org/10.1016/0370-2693(85)90927-X}{\emph{Phys. Lett. B}
  {\bfseries 162} (1985) 299}.

\bibitem{Danielsson:2012et}
U.H.~Danielsson, G.~Shiu, T.~Van~Riet and T.~Wrase, \emph{{A note on obstinate
  tachyons in classical dS solutions}},
  \href{https://doi.org/10.1007/JHEP03(2013)138}{\emph{JHEP} {\bfseries 03}
  (2013) 138} [\href{https://arxiv.org/abs/1212.5178}{{\ttfamily 1212.5178}}].

\bibitem{Andriot:2021rdy}
D.~Andriot, \emph{{Tachyonic de Sitter Solutions of 10d Type II
  Supergravities}},
  \href{https://doi.org/10.1002/prop.202100063}{\emph{Fortsch. Phys.}
  {\bfseries 69} (2021) 2100063}
  [\href{https://arxiv.org/abs/2101.06251}{{\ttfamily 2101.06251}}].

\bibitem{Andriot:2020wpp}
D.~Andriot, P.~Marconnet and T.~Wrase, \emph{{New de Sitter solutions of 10d
  type IIB supergravity}},
  \href{https://doi.org/10.1007/JHEP08(2020)076}{\emph{JHEP} {\bfseries 08}
  (2020) 076} [\href{https://arxiv.org/abs/2005.12930}{{\ttfamily
  2005.12930}}].

\bibitem{Lambers}
J.~Lambers, \emph{{The Method of Steepest descent}}, \href{https://www.math.usm.edu/lambers/mat419/lecture10.pdf}{Lecture 10 Notes MAT 419/519}, (2011).

\bibitem{Achucarro:2018vey}
A.~Ach\'ucarro and G.~A.~Palma, \emph{{The string swampland constraints require multi-field inflation}},   \href{https://doi.org/10.1088/1475-7516/2019/02/041}{\emph{JCAP} {\bfseries 02}
  (2019) 041} [\href{https://arxiv.org/abs/1807.04390}{{\ttfamily
  1807.04390}}].

\bibitem{Hertzberg:2007wc}
M.P.~Hertzberg, S.~Kachru, W.~Taylor and M.~Tegmark, \emph{{Inflationary
  Constraints on Type IIA String Theory}},
  \href{https://doi.org/10.1088/1126-6708/2007/12/095}{\emph{JHEP} {\bfseries
  12} (2007) 095} [\href{https://arxiv.org/abs/0711.2512}{{\ttfamily
  0711.2512}}].




\end{thebibliography}
\end{document}